\documentclass[printer]{aa}
\usepackage{graphicx, subfigure}
\usepackage[graphicx]{realboxes}
\usepackage{appendix}
\usepackage{xcolor, soul}
\sethlcolor{yellow}
\usepackage{txfonts}
\usepackage{supertabular}
\usepackage{eqnarray}
\usepackage{adjustbox}
\usepackage{enumitem}
\usepackage{natbib}
\usepackage{amsmath} 
\usepackage{hyperref} 
\usepackage{longtable}

\definecolor{myblue}{RGB}{0, 160, 240}

\definecolor{myblue}{RGB}{0, 160, 240} 
\definecolor{mygreen}{RGB}{0, 180, 0}

\usepackage{hyperref}
\hypersetup{colorlinks = true, citecolor = {blue}, linkcolor={blue}, urlcolor={blue}}

\begin{document} 

\title{Inspecting the Cepheid parallax of pulsation using \textit{Gaia} EDR3 parallaxes}
\subtitle{Projection factor and period-luminosity and period-radius relations}

\author{B. Trahin \inst{1,2}, 
L. Breuval \inst{1}, 
P. Kervella \inst{1}, 
A. M\'erand \inst{3},  
N. Nardetto \inst{4},
A. Gallenne \inst{5,6,7},
V. Hocd\'e \inst{5},
W. Gieren \inst{6}}

\institute{LESIA, Observatoire de Paris, Universit\'e PSL, CNRS, Sorbonne Universit\'e, Universit\'e de Paris, 5 place Jules Janssen, 92195 Meudon, France
        \and Universit\'e Paris-Saclay, CNRS, Institut d’Astrophysique Spatiale, 91405 Orsay, France\\
        \email{boris.trahin@ias.u-psud.fr}
        \and European Southern Observatory, Karl-Schwarzschild-Str. 2, 85748 Garching, Germany
        \and Universit\'e C\^ote d'Azur, Observatoire de la C\^ote d'Azur, CNRS, Laboratoire Lagrange, France
        \and Nicolaus Copernicus Astronomical Center of the Polish Academy of Sciences, ul. Bartycka 18, PL-00-716 Warszawa, Poland
        \and Universidad de Concepción, Departamento de Astronomía, Casilla 160-C, Concepción, Chile
        \and Unidad Mixta Internacional Franco-Chilena de Astronomia (CNRS UMI 3386), Departamento de Astronomía, Universidad de Chile, Camino el Observatorio 1515, Las Condes, Santiago, Chile
        }

\date{}

\abstract
        {As primary anchors of the distance scale, Cepheid stars play a crucial role in our understanding of the distance scale of the Universe because of their period-luminosity relation. Determining precise and consistent parameters (radius, temperature, color excess, and projection factor) of Cepheid pulsating stars is therefore very important.}
        {With the high-precision parallaxes delivered by the early third \textit{Gaia} data release (EDR3), we aim to derive various parameters of Cepheid stars in order to calibrate the period-luminosity and period-radius relations and to investigate the relation of period to $p$-factor. }
        {We applied an implementation of the parallax-of-pulsation method through the algorithm called spectro-photo-interferometry of pulsating stars (SPIPS), which combines all types of available data for a variable star (multiband and multicolor photometry, radial velocity, effective temperature, and interferometry measurements) in a global modeling of its pulsation. }
        {We present the SPIPS modeling of a sample of 63 Galactic Cepheids. Adopting \textit{Gaia} EDR3 parallaxes as an input associated with the best available dataset, we derive consistent values of parameters for these stars such as the radius, multiband apparent magnitudes, effective temperatures, color excesses, period changes, Fourier parameters, and the projection factor.}
        {Using the best set of data and the most precise distances for Milky Way Cepheids, we derive new calibrations of the period-luminosity and period-radius relations: $M_{K_S} = -5.529_{\pm0.015}-3.141_{\pm0.050}(\log P - 0.9)$ and $\log R = 1.763_{\pm0.003}+0.653_{\pm0.012}(\log P - 0.9)$. After investigating the dependences of the projection factor on the parameters of the stars, we find a high dispersion of its values and no evidence of its correlation with the period or with any other parameters such as radial velocity, temperature, or metallicity. Statistically, the $p-$factor has an average value of $p = 1.26 \pm 0.07$, but with an unsatisfactory agreement ($\sigma = 0.15$). In absence of any clear correlation between the $p-$factor and other quantities, the best agreement is obtained under the assumption that the $p-$factor  can take any value in a band with a width of 0.15. This result highlights the need for a further examination of the physics behind the $p-$factor.}
        

\titlerunning{Modeling the Cepheid pulsation with \textit{Gaia} EDR3 parallaxes}
\authorrunning{Trahin et al.}
\maketitle


\section{Introduction}
\label{sec:intro}

Cepheids are the best-established standard candle. They link the distance scale in the Local Group with type Ia supernova host galaxies. A thorough understanding of the pulsation of these stars is required to obtain the best accuracy on the Hubble constant $H_{0}$ \citep{Breuval2020, Riess2021}. 

Obtaining accurate distances to Cepheid stars is still a nontrivial issue. Cepheid distances may be derived through main-sequence fitting for Cepheids in clusters or through the measurement of their parallax. Recently, very precise geometric parallaxes for about 9500 Cepheids were measured by the \textit{Gaia} satellite \citep{GaiaEDR3}, which is the first competitive alternative to \textit{Hubble} Space Telescope (HST) parallaxes \citep{Benedict2007, Riess2018a}. 

In addition, distances to classical Cepheids (CCs) can be obtained from the parallax-of-pulsation method (PoP). In this approach, the variation in the angular diameter of a Cepheid is compared with the variation of its linear diameter, derived from the integration of its pulsation velocity. The true pulsational velocity of a star is derived by multiplying the disk-integrated radial velocities (measured by spectroscopy) by a projection factor (hereafter $p-$factor). In the absence of interferometric measurements, angular diameters can be derived from surface-brightness-color relations (SBCR): this particular implementation of the PoP technique is known as the Baade-Wesselink (BW) method \citep{Baade1926, Wesselink1946}. The PoP method is the most geometrical way, except for measuring the direct parallax, to estimate the distance of Cepheids. This method is therefore valuable in calibrating the period-luminosity (P$-$L) relation, also called the Leavitt law \citep{Leavitt1912}. However, the assumptions behind the PoP method may introduce strong sources of error on the derived distances. Especially the current uncertainty on the $p-$factor value is still the main reason for recent determinations of the Hubble constant based on Cepheid distances to avoid relying on the PoP technique \citep{Riess2009}.

\citet{Merand2015} developed the code called spectro-photo-interferometry of pulsating stars (SPIPS). This is a variant implementation of the PoP method that uses atmospheric models and combines all types of available data in order to bypass the limitations of the traditional BW method that affect the accuracy and precision of the derived parameters of a pulsating star. Unfortunately, previous studies using this method \citep{Merand2015, Breitfelder2016, Kervella2017, Gallenne2017, Trahin2019} or alternatives \citep{Ngeow2012, Storm2011a, Pilecki2018} did not converge to a consistent dependence of the $p-$factor because the few available HST parallaxes were not very precise, because of the \textit{Gaia} DR2 zeropoint uncertainty, or because the datasets were incomplete. 

In this paper, we present the application of the SPIPS method to a sample of CCs for which we used the best and most complete data, in combination with the new \textit{Gaia} EDR3 parallaxes, and we derive various precise and consistent parameters and investigate their dependences. This paper is similar to the study by \citet{Gallenne2017}, who performed a SPIPS analysis of Large Magellanic Cloud (LMC) and Small Magellanic Cloud (SMC) Cepheids for which they disposed of light curves in order to derive the period-$p-$factor relation. The difference is that our work is based on Milky Way Cepheids and uses a larger set of data (effective temperatures, more complete photometry, and radial velocities and diameters).

In Sect. \ref{sec:data} we introduce our sample of 63 Galactic Cepheids with their data and present the SPIPS method. In Sect. \ref{sec:results} we adopt \textit{Gaia} EDR3 parallaxes as an input into the SPIPS algorithm and apply this method to our sample of Cepheids. Our calculations converge to a robust estimate of their parameters such as radius, reddening, mean multiband magnitudes, effective temperature, and $p-$factor. Finally, in Sect. \ref{sec:discussion} we test the accuracy of the parameters derived from the SPIPS modeling by calibrating the P$-$L and period-radius (P$-$R) relations, and we investigate the dependences of the projection factor.

\section{Cepheid data and fitting method}
\label{sec:data}

        \subsection{Cepheid sample and data}

We built a database including most of the observations collected in the past 50 years for more than 300 Cepheids (including our own observations) in order to identify the stars with the best dataset. The realization of the resulting database was made possible using the McMaster\footnote{McMaster: \url{https://crocus.physics.mcmaster.ca/Cepheid/}}, Vizier \citep{Vizier}, Simbad \citep{Simbad}, AAVSO\footnote{AAVSO: \url{https://www.aavso.org}} , and ADS\footnote{ADS: \url{https://ui.adsabs.harvard.edu}} databases. For the application of the PoP technique, we only used a subset of this database for which the data were ideal. We assumed that a suitable dataset corresponds to a full phase-coverage, which is associated with a good accuracy and a minimum dispersion of the data. Moreover, we required that all data had the corresponding epoch of observation. The Modified Julian Date (MJD) of an observation allowed us to determine the period and the period changes of the star with the best precision, which is not possible with the indication of the phase alone. This preliminary selection led to a sample of 63 Cepheids that covers a broad range of periods from 3 to 68 days. This sample constitutes one of the most complete, precise, and homogeneous samples of Galactic Cepheids that are available for the application of the PoP method. The references of all the data we used are provided in Table \ref{tab:refs} in the appendix.\\

\textbf{Photometry:} Each Cepheid of our sample has at least photometric data in the optical $B$ and $V$ bands (which contain the information about the temperature and reddening), and in the near-infrared (NIR) $J$, $H$, $K$ bands (which are less sensitive to interstellar reddening and are more sensitive to the variation in radius). We also used photometric data from spatial observatories and surveys such as \textit{Hipparcos} ($H_p$ band) and \textit{Tycho} ($B$ and $V$ bands), \textit{Spitzer} ($I_1$ and $I_2$ bands), \textit{2MASS} ($J$, $H,$ and $K_S$ bands), and \textit{Gaia} ($G$, $BP,$ and $RP$ bands). As recommended in \citet{Breitfelder2016}, we did not use $R$- and $I$-band photometry because the effective bandpasses are poorly defined.
In the SPIPS algorithm, all photometric observations are  modeled with the dedicated filters available in the Spanish Virtual Observatory database\footnote{SVO: \url{http://svo2.cab.inta-csic.es/theory/fps3/index.php?mode=browse}} \citep[SVO,][]{Rodrigo2020}. Most of the data used in this study are originally in the California Institute of Technology (CIT) system. However, only the South African Astronomical Observatory (SAAO) filters are not available in the SVO database. We converted the infrared photometry from the SAAO system into the CIT system in order to include these data and to obtain a better phase coverage of the NIR photometry. We used the following equations from \cite{Carter1990}:
\[
\begin{array}{l l l l l l l l}
J_{\rm CIT} ~= J_{\rm SAAO} ~- 0.134~ (J-K)_{\rm SAAO} - 0.001 ~(\sigma=0.010),  \\
H_{\rm CIT} = H_{\rm SAAO} - 0.022 ~(J-K)_{\rm SAAO} + 0.004~(\sigma=0.013),   \\
K_{\rm CIT} = K_{\rm SAAO} ~- 0.027 ~(J-K)_{\rm SAAO} - 0.003~(\sigma=0.010).   \\
\end{array}
\]
In section \ref{sec:PL} we perform a second transformation of mean magnitudes from the CIT to the 2MASS system in order to compare our PL relations in infrared bands with other calibrations from the literature. In the SPIPS adjustments, we decided to keep the data (except for the SAAO data) in their original system as far as possible in order to avoid introducing potential systematics in the derived parameters.
For safety, we introduced a conservative systematic uncertainty of 0.01 mag in order to take the different instrumental calibrations and photometric zeropoints into account. This value is consistent with the average offset that is generally observed when data from different instruments and magnitude systems are combined \citep[see, e.g.,][]{Barnes1997,Breitfelder2016}. \\

\textbf{Radial velocities:} The $p$-factor depends on the method that is used to extract the radial velocity (such as cross-correlation or broadening functions) because the velocity curves that are obtained with different techniques can have a difference of up to 5\% in amplitude \citep{Nardetto2009}. This must be taken into account for studies that use the $p$-factor, in particular regarding its dependence on other parameters such as the period. In this work, we only used radial velocities determined from cross-correlation techniques. As the $p$-factor directly depends on the integrated radial velocity curve, we took care to use only precise observations with full phase coverage and with a well-defined amplitude. As observed in \cite{Kervella2019a}, at least 80\% of the Cepheids belong to a multiple system. For most stars of our sample, binary Cepheids are not excluded, but the effect on the radial velocities and photometry is considered to be negligible. For some Cepheids, radial velocities are clearly affected by a spectroscopic companion and were corrected for the Keplerian motion using the orbital parameters from the Konkoly database\footnote{Konkoly orbital parameters: \url{https://konkoly.hu/CEP/orbit.html}} in order to retain only the pulsation component. They are indicated by a star in Table \ref{tab:refs}. 

A conservative uncertainty of 0.5 km.s$^{-1}$ was quadratically included as a systematic error in order to take all the systematic effects due to the combination of different datasets into account.
\\

\textbf{Effective temperatures:} For some stars, we disposed of effective temperature measurements, which are mostly provided by the series of papers by \citet{Luck2004}, \citet{Andrievsky2005}, \citet{Kovtyukh2005}, \citet{Luck2008}, \citet{Luck2018}, and \citet{Proxauf2018}. In these papers, the authors estimated the depth ratio of about 50 spectral lines \citep[described in][]{Kovtyukh1999} in order to derive the effective temperature of the star. These observations allowed us to constrain the SPIPS models better and to evaluate the consistency of the atmospheric models.
We included an error of 50 K as a systematic error for the effective temperatures \citep{Breitfelder2016}.
\\

\textbf{Angular diameters:} In the past ten years, improvements in interferometry enabled the direct determination of the angular diameter for some Cepheids. Several stars of this sample were regularly observed with the CHARA and VLTI interferometers in order to obtain direct measurements of their angular diameter variations. These observations associated with the SPIPS method already allowed us to obtain a better precision on the projection factor \citep{Breitfelder2016}. The new raw data that we obtained with the PIONIER instrument of the VLTI were reduced using the \texttt{pndrs} data reduction software \citep{LeBouquin2011}. We then adjusted the calibrated squared visibilities with a uniform disk (UD) model to obtain the UD angular diameters.
A conservative uncertainty corresponding to 2\% of the angular diameter values was quadratically added as a systematic \citep{Kervella2004a}.\\

\begin{figure*}[]
\centering
\includegraphics[width=\textwidth]{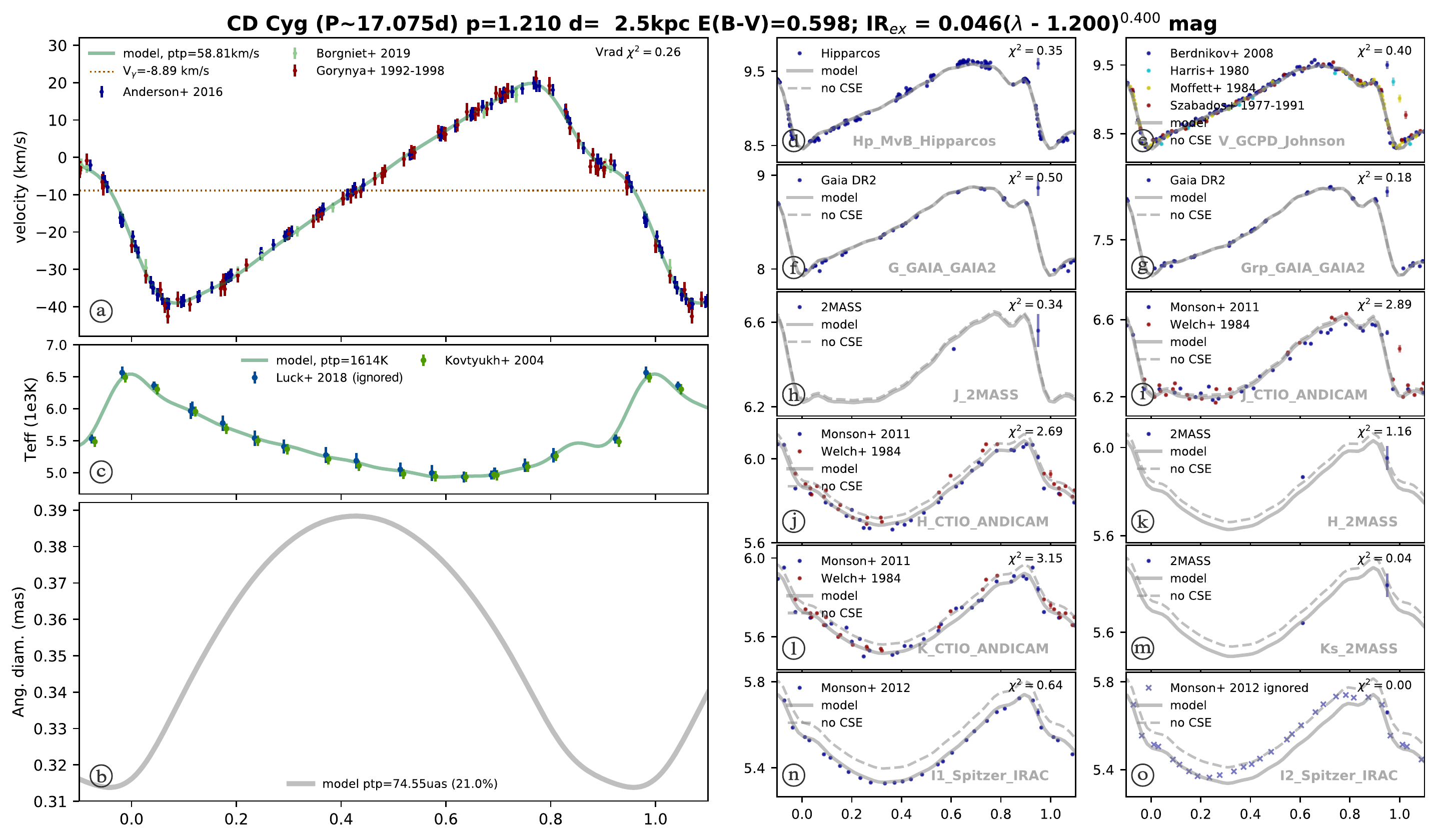}
\caption{Result of the SPIPS modeling for the Cepheid CD Cyg. The fitted observed data include radial velocities (top left), effective temperatures (middle left), and multiband photometry (right). Some main parameters derived from this modeling are listed above the plot. The distance is fixed to the \textit{Gaia} EDR3 parallax with the zeropoint correction by \cite{Lindegren2020b}.}
\label{fig:CD_Cyg}
\end{figure*}

\textbf{Distances:} As input in the SPIPS code, we adopted the parallaxes from \textit{Gaia} EDR3 \citep{GaiaEDR3} and inverted them to obtain the Cepheid distances. We note that using the \cite{BailerJones2021} approach to derive distances of 9,000 Cepheids, the geometric distance (based on the parallax and on the direction on the sky) and the photo-geometric distance (which also includes the color and apparent magnitude of the star) differ by 9 pc from the inverted \textit{Gaia} EDR3 parallaxes on average, with a largest difference of 230 pc. This comparison confirms that the inversion of \textit{Gaia} EDR3 parallaxes in order to obtain Cepheid distances does not add biases to the computed parameters.
We corrected each parallax for their individual zeropoint offset by using the dedicated Python code\footnote{EDR3 zeropoint code: \url{https://www.cosmos.esa.int/web/Gaia/edr3-code}} described by \citet{Lindegren2020a}. Alternative procedures to determine the zeropoint offsets were realized \cite[][e.g., suggested that the \textit{Gaia} EDR3 parallaxes may be underestimated by about 5\%]{Groenewegen2021}, but we limit this study to the \cite{Lindegren2020a} corrections. The new \textit{Gaia} EDR3 catalog also provides the renormalized unit weight error (RUWE) indicator, which represents the quality of a star's parallax compared with other stars of the same type. \citet{Lindegren2020a} recommended to avoid the use of parallaxes with a RUWE indicator higher than 1.4. We find 16 stars in this case in our sample of 63 Cepheids: we performed the SPIPS modeling successfully for these stars, but we did not use them to calibrate the P$-p$, P$-$L, and P$-$R relations, which depend on the distance. However, we made an exception for $\delta$ Cep, for which we had one of the best available datasets, with a full coverage of the interferometric angular diameters and spectroscopic effective temperature curves. For this star, \cite{Kervella2019b} found a close companion that has a very precise \textit{Gaia} EDR3 parallax with a RUWE of 1.415, which is only slightly higher than the threshold for the other stars of our sample. Only one other star (RS Pup) has a similar dataset, which permits constraining the different parameters better.

The range of magnitudes $G = [10.8-11.2]$ corresponds to a transition of window classes \citep[see Fig. 1 in][]{Lindegren2020b} that might affect the accuracy of the zeropoint offset, but none of our stars falls in this range. Finally, we followed the conservative recommendation by \citet{Riess2021} and increased each parallax error by 10 \% to account for potential additional excess uncertainty.

        \subsection{SPIPS fitting method}
        \label{sec:spips}
We used the SPIPS modeling tool\footnote{The SPIPS algorithm is available at: \url{https://github.com/amerand/SPIPS}} from \citet{Merand2015} to reproduce our observational dataset. This algorithm is inspired by the classical BW technique. We here present the general idea of the SPIPS method and refer the reader to \citet{Merand2015} for more details.

The motivation behind the SPIPS method is to bypass the limitations of the traditional BW implementation, which affect the accuracy and precision of the derived parameters. A main limitation of the BW method results from the determination of angular diameters through surface brightness-color relations using only two photometric bands (generally $V$ and $K$). In this case, the effective temperature and the angular diameter of the star are adjusted from only two photometric measurements. Finally, a poor phase coverage or a low-order interpolation of the different quantities can prevent the precise determination of the parameters.

The approach of the SPIPS method is first to propose a combination of all the data available in the literature for a star. This includes spectroscopic radial velocities as well as photometric measurements in any filter and optical interferometric measurements. In the current code, we use radial velocities derived from cross-correlation. A future implementation is in progress to directly reproduce the spectral lines from high-resolution spectroscopy to derive RVs, effective temperatures, and other parameters.
The data are then adjusted simultaneously altogether, using a standard multiparameter $\chi^2$ minimization, in order to obtain more realistic estimates of the statistical uncertainties, as opposed to a method that would fit consecutive sets of parameters. The SPIPS code also determines the period as well as the period changes of the pulsation by phasing the data. 
The BW method generally makes the assumption that empirical surface-brightness relations are linear in color (e.g., $V-K$), which propagates a color bias on the distance. In order to bypass these uncertainties, SPIPS computes the specific surface brightness using a grid of ATLAS9 atmospheric models\footnote{ATLAS9 atmospheric models are available on: \url{http://wwwuser.oats.inaf.it/castelli/grids.html}} \citep{Castelli2004} to derive synthetic photometry from the effective temperature. The photometric magnitudes are then computed on this grid, using bandpasses and zeropoints from the SVO database. If interferometric observations of the angular diameter of a star are available, the effects of the limb darkening have to be taken into account: in the SPIPS algorithm, the uniform disk angular diameters estimated from the observed visibilities are converted into limb-darkening values using SATLAS\footnote{SATLAS: \url{http://cdsarc.u-strasbg.fr/viz-bin/qcat?J/A+A/554/A98}} spherical atmosphere models \citep{Neilson2013}.

The interstellar reddening is parameterized in SPIPS using the $B-V$ color excess $E(B-V)$ and the reddening law from \citet{Fitzpatrick1999}, adopting $R_V=3.1$. As explained in \citet{Merand2015}, the reddening corrections in SPIPS are computed on the basis of photometric observations of the star, whereas in classical implementations of the BW method, they are usually computed  for a Vega-like star, which is much hotter (10000 K) than Cepheids ($\sim 5000$ K). Moreover, a circumstellar envelope (CSE) is a frequent phenomenon around massive pulsating stars such as Cepheids \citep{Hocde2020, Gallenne2021}. It introduces a bias on the interferometric angular diameters and the NIR photometric measurements.
The latter are characterized by a magnitude excess and are taken into account in SPIPS by adjusting a power law for the infrared excess, assuming that there is no excess in optical wavelengths ($\lambda<1.2$ $\mu$m). This law is defined as IR$_{\mathrm{ex}} = \alpha(\lambda-1.2)^{0.4}$ , where $\alpha$ is the slope of the relation and $\lambda$ is the wavelength of the photometric band in $\mu$m. Regarding interferometric measurements, the bias due to a CSE depends on the baselines and on the angular diameter. It is tabulated in SPIPS as a function of the infrared excess. The visibilities of a limb-darkened disk surrounded by a CSE are synthesized, and a uniform disk model is then adjusted to estimate the bias.

For each CC, the phases of the data points were calculated using their corresponding modified Julian date epoch of maximum light (MJD$_0$). A strategic approach to fitting the data was to start from a model whose general properties were close to the observed data so that the model fitting would converge faster. These starting model curves are third-order Fourier series whose amplitudes and phases agree with the data. They were built from a set of parameters found in the literature (e.g., the mean effective temperature, the period, and the MJD$_0$) and by computing mean values of the radial velocity and of the angular diameter from the available data and distance of the star. Depending on the properties of the different curves (e.g., bumps or steep variations), we then adapted the number of Fourier modes and thus of free parameters to obtain a satisfactory representation of the observed variations of the star.
The Fourier series decomposition is a robust method for studying the light curves of variable stars. Fourier coefficients and parameters are commonly used today to model a Cepheid light curve \citep{Morgan2007, Deb2009, Bhardwaj2015}. The third-order coefficients in the $K$ band are listed for each star in \ref{tab:fourier}. These results can be a used in future comparative studies aiming at constraining theoretical stellar pulsation models and determining pulsation modes of Cepheids.

\newpage
\setlength{\tabcolsep}{7pt}
\begin{table*}[h!]
\caption{Best-fit mean parameters derived from the SPIPS modeling of our Cepheid sample.}
\footnotesize
\centering
\rotatebox{90}{
\begin{tabular}{l c c c c c c c c c c c c}
\hline
\hline
Star & Plx$_{\rm EDR3}^\star$ & RUWE$^\star$ & MJD$_0^\star$ & Period & $p$-factor & $\Theta$ & $R$ & $E(B-V)$ & T$_{\rm eff}$ & $v_{\gamma}$ & IR$_{\rm ex}$ & $[$Fe/H$]^\star$\\
 & (mas) &  & (days)& (days) &  & (mas) & (R$_{\odot}$) & (mag) & (K) & (km.s$^{-1}$) & (mag) & (dex)\\
\hline \\
V1162 Aql       &       0.823$_{\pm 0.023}$&  0.95& 25802.8229&5.376$_{\pm 6.10^{-06}}$ &1.404$_{\pm 0.072}$ &     0.325$_{\pm 0.007}$     &42.5$_{\pm 1.9}$&  0.225$_{\pm 0.021}$&    5693$_{\pm 80}$ &       13.06$_{\pm 0.51}$      &       0.00$_{\pm 0.02}$ & 0.01$_{\pm 0.08}$\\ 
TT Aql  &       0.997$_{\pm 0.023}$&  1.08& 48308.5708&13.755$_{\pm 3.10^{-05}}$ &1.218$_{\pm 0.032}$ &  0.760$_{\pm 0.015}$     &82.0$_{\pm 3.1}$&      0.551$_{\pm 0.021}$&        5470$_{\pm 80}$ &       3.72$_{\pm 0.51}$       &       0.01$_{\pm 0.02}$ & 0.22$_{\pm 0.06}$\\ 
U Aql   &       1.765$_{\pm 0.088}$&  3.09& 34922.0864&7.024$_{\pm 2.10^{-05}}$ &1.219$_{\pm 0.067}$ &  0.778$_{\pm 0.016}$     &47.4$_{\pm 3.5}$&      0.417$_{\pm 0.021}$&        5735$_{\pm 80}$ &       1.53$_{\pm 0.51}$       &       0.01$_{\pm 0.02}$ & 0.17$_{\pm 0.06}$\\ 
FM Aql  &       1.014$_{\pm 0.026}$&  1.26& 35151.2026&6.114$_{\pm 7.10^{-06}}$ &1.064$_{\pm 0.039}$ &  0.465$_{\pm 0.009}$     &49.3$_{\pm 2.0}$&      0.669$_{\pm 0.021}$&        5788$_{\pm 80}$ &       -5.32$_{\pm 0.51}$      &       0.04$_{\pm 0.02}$ & 0.24$_{\pm 0.06}$\\ 
SZ Aql  &       0.525$_{\pm 0.020}$&  0.94& 54228.3329&17.142$_{\pm 6.10^{-05}}$ &1.124$_{\pm 0.047}$ &  0.464$_{\pm 0.009}$     &94.9$_{\pm 5.5}$&      0.692$_{\pm 0.021}$&        5463$_{\pm 80}$ &       7.78$_{\pm 0.53}$       &       0.01$_{\pm 0.02}$ & 0.18$_{\pm 0.08}$\\ 
FN Aql  &       0.736$_{\pm 0.025}$&  1.12& 36803.2777&9.483$_{\pm 4.10^{-05}}$ &1.273$_{\pm 0.050}$ &  0.407$_{\pm 0.008}$     &59.4$_{\pm 3.1}$&      0.531$_{\pm 0.021}$&        5547$_{\pm 80}$ &       12.97$_{\pm 0.50}$      &       0.03$_{\pm 0.02}$ & -0.06$_{\pm 0.06}$\\ 
$\eta$ Aql      &       3.711$_{\pm 0.195}$&  2.56& 48069.3905&7.177$_{\pm 1.10^{-05}}$ &1.115$_{\pm 0.063}$ &     1.728$_{\pm 0.035}$     &50.1$_{\pm 3.9}$&  0.162$_{\pm 0.021}$&    5790$_{\pm 80}$ &       -14.77$_{\pm 0.52}$     &       0.07$_{\pm 0.02}$ & 0.08$_{\pm 0.08}$\\ 
SY Aur  &       0.462$_{\pm 0.020}$&  1.08& 36843.2737&10.144$_{\pm 6.10^{-05}}$ &1.101$_{\pm 0.058}$ &  0.237$_{\pm 0.005}$     &55.1$_{\pm 3.6}$&      0.510$_{\pm 0.021}$&        6005$_{\pm 80}$ &       -3.19$_{\pm 0.50}$      &       0.07$_{\pm 0.02}$ & -0.07$_{\pm 0.15}$\\ 
RT Aur  &       1.858$_{\pm 0.123}$&  6.43& 47956.9052&3.728$_{\pm 5.10^{-06}}$ &1.525$_{\pm 0.119}$ &  0.703$_{\pm 0.014}$     &40.7$_{\pm 3.9}$&      0.096$_{\pm 0.021}$&        6008$_{\pm 80}$ &       19.68$_{\pm 1.14}$      &       0.01$_{\pm 0.02}$ & 0.13$_{\pm 0.06}$\\ 
VY Car  &       0.565$_{\pm 0.017}$&  0.92& 48339.2975&18.902$_{\pm 4.10^{-05}}$ &1.147$_{\pm 0.037}$ &  0.517$_{\pm 0.010}$     &98.4$_{\pm 4.6}$&      0.347$_{\pm 0.021}$&        5370$_{\pm 80}$ &       1.69$_{\pm 0.51}$       &       0.04$_{\pm 0.02}$ & -0.06$_{\pm 0.15}$\\ 
$\ell$ Car      &       1.988$_{\pm 0.111}$&  2.39& 47774.2368&35.552$_{\pm 1.10^{-03}}$ &1.221$_{\pm 0.068}$ &     2.912$_{\pm 0.058}$     &157.5$_{\pm 12.8}$& 0.223$_{\pm 0.021}$&    5021$_{\pm 80}$ &       3.64$_{\pm 0.50}$       &       0.04$_{\pm 0.02}$ & 0.10$_{\pm 0.15}$\\ 
DD Cas  &       0.346$_{\pm 0.013}$&  1.05& 42780.1785&9.811$_{\pm 3.10^{-05}}$ &1.391$_{\pm 0.062}$ &  0.193$_{\pm 0.004}$     &59.9$_{\pm 3.4}$&      0.514$_{\pm 0.021}$&        5612$_{\pm 80}$ &       -69.54$_{\pm 0.51}$     &       0.05$_{\pm 0.02}$ & 0.10$_{\pm 0.08}$\\ 
CF Cas  &       0.316$_{\pm 0.012}$&  1.04& 37021.2586&4.875$_{\pm 6.10^{-06}}$ &1.376$_{\pm 0.062}$ &  0.122$_{\pm 0.002}$     &41.4$_{\pm 2.5}$&      0.623$_{\pm 0.021}$&        5690$_{\pm 80}$ &       -77.91$_{\pm 0.51}$     &       0.02$_{\pm 0.02}$ & 0.02$_{\pm 0.06}$\\ 
SW Cas  &       0.461$_{\pm 0.012}$&  1.12& 42989.0809&5.441$_{\pm 1.10^{-05}}$ &1.320$_{\pm 0.075}$ &  0.196$_{\pm 0.004}$     &45.7$_{\pm 2.0}$&      0.536$_{\pm 0.021}$&        5846$_{\pm 81}$ &       -41.55$_{\pm 0.54}$     &       0.07$_{\pm 0.02}$ & -0.03$_{\pm 0.08}$\\ 
DL Cas  &       0.580$_{\pm 0.028}$&  1.88& 42779.7290&8.001$_{\pm 2.10^{-05}}$ &1.158$_{\pm 0.062}$ &  0.306$_{\pm 0.006}$     &56.8$_{\pm 4.0}$&      0.568$_{\pm 0.021}$&        5690$_{\pm 80}$ &       -31.90$_{\pm 0.50}$     &       0.02$_{\pm 0.02}$ & -0.01$_{\pm 0.08}$\\ 
KN Cen  &       0.251$_{\pm 0.018}$&  1.03& 54345.3702&34.019$_{\pm 2.10^{-04}}$ &1.315$_{\pm 0.099}$ &  0.423$_{\pm 0.009}$     &181.5$_{\pm 19.3}$&    0.874$_{\pm 0.023}$&        5036$_{\pm 86}$ &       -41.72$_{\pm 0.69}$     &       0.06$_{\pm 0.02}$ & 0.55$_{\pm 0.12}$\\ 
$\delta$ Cep    &       3.578$_{\pm 0.148}$&  2.71& 36075.0094&5.366$_{\pm 2.10^{-06}}$ &1.317$_{\pm 0.056}$ &     1.454$_{\pm 0.029}$     &43.7$_{\pm 2.7}$&  0.092$_{\pm 0.021}$&    5907$_{\pm 80}$ &       -18.21$_{\pm 0.50}$     &       0.06$_{\pm 0.02}$ & 0.12$_{\pm 0.06}$\\ 
$\delta$ Cep$^{\rm comp}$       &       3.483$_{\pm 0.051}$&  1.42& 36075.0112&5.366$_{\pm 2.10^{-06}}$ &1.353$_{\pm 0.025}$ &     1.454$_{\pm 0.029}$     &44.9$_{\pm 1.3}$&  0.092$_{\pm 0.021}$&    5907$_{\pm 80}$ &       -18.21$_{\pm 0.50}$     &       0.06$_{\pm 0.02}$ & 0.12$_{\pm 0.06}$\\ 
V0459 Cyg       &       0.382$_{\pm 0.014}$&  1.09& 36807.8044&7.252$_{\pm 2.10^{-05}}$ &1.534$_{\pm 0.089}$ &     0.207$_{\pm 0.004}$     &58.3$_{\pm 3.2}$&  0.801$_{\pm 0.021}$&    5640$_{\pm 81}$ &       -22.92$_{\pm 0.55}$     &       0.04$_{\pm 0.02}$ & 0.09$_{\pm 0.06}$\\ 
SZ Cyg  &       0.445$_{\pm 0.012}$&  0.96& 43306.9551&15.109$_{\pm 1.10^{-04}}$ &1.183$_{\pm 0.041}$ &  0.348$_{\pm 0.007}$     &84.1$_{\pm 3.7}$&      0.691$_{\pm 0.021}$&        5323$_{\pm 80}$ &       -9.95$_{\pm 0.52}$      &       0.07$_{\pm 0.02}$ & 0.09$_{\pm 0.08}$\\ 
V0538 Cyg       &       0.394$_{\pm 0.017}$&  0.99& 42772.4485&6.119$_{\pm 3.10^{-05}}$ &1.429$_{\pm 0.105}$ &     0.176$_{\pm 0.004}$     &48.1$_{\pm 3.2}$&  0.668$_{\pm 0.021}$&    5714$_{\pm 80}$ &       -17.31$_{\pm 0.60}$     &       0.07$_{\pm 0.02}$ & 0.05$_{\pm 0.06}$\\ 
V0402 Cyg       &       0.410$_{\pm 0.011}$&  0.92& 41698.0525&4.365$_{\pm 7.10^{-06}}$ &1.403$_{\pm 0.067}$ &     0.160$_{\pm 0.003}$     &42.0$_{\pm 1.8}$&  0.416$_{\pm 0.021}$&    5761$_{\pm 80}$ &       -13.45$_{\pm 0.52}$     &       0.03$_{\pm 0.02}$ & 0.02$_{\pm 0.08}$\\ 
CD Cyg  &       0.394$_{\pm 0.016}$&  1.01& 48321.6400&17.075$_{\pm 6.10^{-05}}$ &1.210$_{\pm 0.054}$ &  0.355$_{\pm 0.007}$     &97.0$_{\pm 6.1}$&      0.598$_{\pm 0.021}$&        5445$_{\pm 80}$ &       -8.89$_{\pm 0.52}$      &       0.05$_{\pm 0.02}$ & 0.15$_{\pm 0.06}$\\ 
X Cyg   &       0.910$_{\pm 0.020}$&  1.28& 48319.5377&16.386$_{\pm 3.10^{-05}}$ &1.258$_{\pm 0.032}$ &  0.811$_{\pm 0.016}$     &95.9$_{\pm 3.6}$&      0.297$_{\pm 0.021}$&        5315$_{\pm 80}$ &       7.66$_{\pm 0.50}$       &       0.03$_{\pm 0.02}$ & 0.10$_{\pm 0.08}$\\ 
MW Cyg  &       0.542$_{\pm 0.019}$&  1.21& 42923.4094&5.955$_{\pm 1.10^{-05}}$ &1.614$_{\pm 0.080}$ &  0.300$_{\pm 0.006}$     &59.6$_{\pm 3.3}$&      0.754$_{\pm 0.021}$&        5793$_{\pm 80}$ &       -12.49$_{\pm 0.51}$     &       0.04$_{\pm 0.02}$ & 0.09$_{\pm 0.08}$\\ 
V0386 Cyg       &       0.894$_{\pm 0.013}$&  0.95& 42776.4566&5.258$_{\pm 2.10^{-05}}$ &1.371$_{\pm 0.078}$ &     0.374$_{\pm 0.008}$     &45.1$_{\pm 1.3}$&  0.929$_{\pm 0.022}$&    5739$_{\pm 82}$ &       -6.73$_{\pm 0.54}$      &       0.09$_{\pm 0.02}$ & 0.11$_{\pm 0.08}$\\ 
VZ Cyg  &       0.545$_{\pm 0.016}$&  1.31& 41705.1890&4.864$_{\pm 5.10^{-06}}$ &1.248$_{\pm 0.049}$ &  0.199$_{\pm 0.004}$     &39.3$_{\pm 1.9}$&      0.287$_{\pm 0.021}$&        5799$_{\pm 80}$ &       -19.93$_{\pm 0.50}$     &       0.04$_{\pm 0.02}$ & 0.05$_{\pm 0.08}$\\ 
$\beta$ Dor     &       2.937$_{\pm 0.140}$&  4.54& 50274.9458&9.843$_{\pm 2.10^{-05}}$ &1.382$_{\pm 0.070}$ &     1.763$_{\pm 0.035}$     &64.6$_{\pm 4.5}$&  0.063$_{\pm 0.021}$&    5581$_{\pm 80}$ &       9.52$_{\pm 0.51}$       &       0.07$_{\pm 0.02}$ & -0.14$_{\pm 0.09}$\\ 
$\zeta$ Gem     &       3.112$_{\pm 0.219}$&  2.78& 48707.9234&10.150$_{\pm 3.10^{-05}}$ &1.165$_{\pm 0.085}$ &     1.637$_{\pm 0.033}$     &56.6$_{\pm 5.7}$&  0.040$_{\pm 0.021}$&    5520$_{\pm 80}$ &       5.79$_{\pm 0.50}$       &       0.06$_{\pm 0.02}$ & -0.19$_{\pm 0.09}$\\ 
V Lac   &       0.496$_{\pm 0.016}$&  1.09& 28900.5590&4.984$_{\pm 3.10^{-06}}$ &1.469$_{\pm 0.072}$ &  0.214$_{\pm 0.004}$     &46.4$_{\pm 2.3}$&      0.430$_{\pm 0.021}$&        6144$_{\pm 80}$ &       -24.04$_{\pm 0.96}$     &       0.06$_{\pm 0.02}$ & 0.06$_{\pm 0.06}$\\ 
BG Lac  &       0.581$_{\pm 0.019}$&  1.43& 35314.3374&5.332$_{\pm 6.10^{-06}}$ &1.147$_{\pm 0.045}$ &  0.220$_{\pm 0.004}$     &40.8$_{\pm 2.0}$&      0.324$_{\pm 0.021}$&        5759$_{\pm 80}$ &       -17.42$_{\pm 0.51}$     &       0.06$_{\pm 0.02}$ & 0.07$_{\pm 0.06}$\\ 
RR Lac  &       0.424$_{\pm 0.015}$&  1.10& 42776.2031&6.416$_{\pm 1.10^{-05}}$ &1.308$_{\pm 0.067}$ &  0.225$_{\pm 0.005}$     &57.0$_{\pm 3.2}$&      0.388$_{\pm 0.021}$&        5930$_{\pm 80}$ &       -38.21$_{\pm 0.51}$     &       0.02$_{\pm 0.02}$ & 0.04$_{\pm 0.06}$\\ 
Z Lac   &       0.510$_{\pm 0.021}$&  1.05& 48313.0699&10.886$_{\pm 3.10^{-05}}$ &1.375$_{\pm 0.065}$ &  0.342$_{\pm 0.007}$     &72.1$_{\pm 4.4}$&      0.460$_{\pm 0.021}$&        5647$_{\pm 80}$ &       -29.52$_{\pm 0.52}$     &       0.03$_{\pm 0.02}$ & 0.10$_{\pm 0.06}$\\ 
Y Lac   &       0.431$_{\pm 0.013}$&  1.05& 41746.2637&4.324$_{\pm 3.10^{-06}}$ &1.168$_{\pm 0.048}$ &  0.154$_{\pm 0.003}$     &38.5$_{\pm 1.8}$&      0.206$_{\pm 0.021}$&        5905$_{\pm 80}$ &       -23.48$_{\pm 0.91}$     &       0.08$_{\pm 0.02}$ & 0.03$_{\pm 0.06}$\\ 
CV Mon  &       0.601$_{\pm 0.015}$&  1.10& 42772.6486&5.379$_{\pm 1.10^{-05}}$ &1.242$_{\pm 0.052}$ &  0.237$_{\pm 0.005}$     &42.4$_{\pm 1.7}$&      0.826$_{\pm 0.021}$&        5718$_{\pm 80}$ &       19.49$_{\pm 0.52}$      &       0.03$_{\pm 0.02}$ & 0.09$_{\pm 0.09}$\\ 
T Mon   &       0.745$_{\pm 0.052}$&  1.72& 43783.9527&27.026$_{\pm 3.10^{-04}}$ &1.146$_{\pm 0.082}$ &  0.944$_{\pm 0.019}$     &136.4$_{\pm 13.7}$&    0.271$_{\pm 0.021}$&        5172$_{\pm 80}$ &       22.03$_{\pm 0.59}$      &       0.03$_{\pm 0.02}$ & -0.04$_{\pm 0.09}$\\ 
S Mus   &       1.179$_{\pm 0.092}$&  4.50& 40300.7624&9.660$_{\pm 3.10^{-05}}$ &1.170$_{\pm 0.101}$ &  0.684$_{\pm 0.014}$     &62.4$_{\pm 7.0}$&      0.287$_{\pm 0.021}$&        5941$_{\pm 80}$ &       -1.53$_{\pm 0.51}$      &       0.07$_{\pm 0.02}$ & 0.13$_{\pm 0.09}$\\ 
S Nor   &       1.099$_{\pm 0.022}$&  0.88& 44018.5575&9.754$_{\pm 3.10^{-05}}$ &1.226$_{\pm 0.037}$ &  0.657$_{\pm 0.013}$     &64.3$_{\pm 2.2}$&      0.290$_{\pm 0.021}$&        5723$_{\pm 80}$ &       5.40$_{\pm 0.51}$       &       0.04$_{\pm 0.02}$ & 0.02$_{\pm 0.09}$\\ 
AW Per  &       1.093$_{\pm 0.029}$&  1.16& 42708.6556&6.464$_{\pm 1.10^{-05}}$ &1.175$_{\pm 0.056}$ &  0.534$_{\pm 0.011}$     &52.6$_{\pm 2.2}$&      0.553$_{\pm 0.021}$&        5957$_{\pm 80}$ &       2.98$_{\pm 0.74}$       &       0.01$_{\pm 0.02}$ & 0.04$_{\pm 0.06}$\\ 
RS Pup  &       0.581$_{\pm 0.017}$&  1.16& 54215.7999&41.454$_{\pm 7.10^{-04}}$ &1.228$_{\pm 0.040}$ &  0.924$_{\pm 0.019}$     &171.0$_{\pm 8.0}$&     0.590$_{\pm 0.021}$&        5284$_{\pm 80}$ &       25.28$_{\pm 0.59}$      &       0.05$_{\pm 0.02}$ & 0.07$_{\pm 0.15}$\\ 
AQ Pup  &       0.294$_{\pm 0.023}$&  1.18& 54587.1360&30.167$_{\pm 3.10^{-04}}$ &1.253$_{\pm 0.099}$ &  0.445$_{\pm 0.009}$     &162.9$_{\pm 18.1}$&    0.528$_{\pm 0.023}$&        5075$_{\pm 84}$ &       60.82$_{\pm 0.52}$      &       -0.00$_{\pm 0.02}$ & 0.06$_{\pm 0.05}$\\  
\hline
\end{tabular}}
\\
\tablefoot{Nonfitted parameters are indicated by a star ($^{\star}$). Systematic uncertainties are taken into account in the listed errors (see also Sect. \ref{sec:data}).}
\label{tab:parameters}
\end{table*}
\newpage

\setlength{\tabcolsep}{7pt}
\begin{table*}[h!]
\caption{Table. \ref{tab:parameters} (continued)}
\footnotesize
\centering
\rotatebox{90}{
\begin{tabular}{l c c c c c c c c c c c c}
\hline
\hline
Star & Plx$_{\rm EDR3}^\star$ & RUWE$^\star$ & MJD$_0^\star$ & Period & $p$-factor & $\Theta$ & $R$ & $E(B-V)$ & T$_{\rm eff}$ & $v_{\gamma}$ & IR$_{\rm ex}$ & $[$Fe/H$]^\star$\\
 & (mas) &  & (days)& (days) &  & (mas) & (R$_{\odot}$) & (mag) & (K) & (km.s$^{-1}$) & (mag)& (dex)\\
\hline \\
VZ Pup  &       0.220$_{\pm 0.015}$&  1.24& 41121.1536&23.174$_{\pm 2.10^{-04}}$ &1.294$_{\pm 0.092}$ &  0.221$_{\pm 0.004}$     &108.2$_{\pm 10.5}$&    0.536$_{\pm 0.022}$&        5595$_{\pm 81}$ &       62.88$_{\pm 0.54}$      &       0.04$_{\pm 0.02}$ & -0.01$_{\pm 0.04}$\\ 
X Pup   &       0.397$_{\pm 0.020}$&  1.04& 54143.6693&25.971$_{\pm 1.10^{-04}}$ &0.762$_{\pm 0.053}$ &  0.381$_{\pm 0.008}$     &103.2$_{\pm 7.8}$&     0.547$_{\pm 0.022}$&        5567$_{\pm 80}$ &       71.34$_{\pm 0.52}$      &       0.11$_{\pm 0.03}$ & 0.02$_{\pm 0.08}$\\ 
LS Pup  &       0.214$_{\pm 0.016}$&  1.25& 38375.6462&14.147$_{\pm 6.10^{-05}}$ &1.144$_{\pm 0.087}$ &  0.156$_{\pm 0.003}$     &78.2$_{\pm 8.3}$&      0.606$_{\pm 0.023}$&        5781$_{\pm 88}$ &       81.01$_{\pm 0.51}$      &       0.12$_{\pm 0.02}$ & -0.12$_{\pm 0.11}$\\ 
RY Sco  &       0.764$_{\pm 0.032}$&  0.73& 54670.5015&20.323$_{\pm 2.10^{-04}}$ &1.222$_{\pm 0.061}$ &  0.778$_{\pm 0.016}$     &109.5$_{\pm 6.9}$&     0.747$_{\pm 0.021}$&        5170$_{\pm 80}$ &       -18.71$_{\pm 1.11}$     &       -0.00$_{\pm 0.02}$ & 0.01$_{\pm 0.06}$\\ 
V0636 Sco       &       1.180$_{\pm 0.034}$&  1.15& 51402.3165&6.797$_{\pm 8.10^{-06}}$ &1.463$_{\pm 0.131}$ &     0.590$_{\pm 0.012}$     &53.8$_{\pm 2.4}$&  0.234$_{\pm 0.021}$&    5575$_{\pm 80}$ &       8.81$_{\pm 0.51}$       &       0.04$_{\pm 0.02}$ & 0.10$_{\pm 0.06}$\\ 
SS Sct  &       0.934$_{\pm 0.023}$&  0.84& 35315.0724&3.671$_{\pm 3.10^{-06}}$ &1.028$_{\pm 0.054}$ &  0.305$_{\pm 0.006}$     &35.1$_{\pm 1.4}$&      0.398$_{\pm 0.021}$&        5973$_{\pm 81}$ &       -7.29$_{\pm 0.52}$      &       0.04$_{\pm 0.02}$ & 0.14$_{\pm 0.06}$\\ 
Z Sct   &       0.357$_{\pm 0.018}$&  0.90& 36246.6385&12.901$_{\pm 5.10^{-05}}$ &1.190$_{\pm 0.071}$ &  0.268$_{\pm 0.005}$     &80.9$_{\pm 6.0}$&      0.642$_{\pm 0.021}$&        5566$_{\pm 81}$ &       30.02$_{\pm 0.51}$      &       0.01$_{\pm 0.02}$ & 0.12$_{\pm 0.09}$\\ 
S Sge   &       1.700$_{\pm 0.111}$&  4.00& 42678.3060&8.382$_{\pm 2.10^{-05}}$ &0.980$_{\pm 0.067}$ &  0.778$_{\pm 0.016}$     &49.2$_{\pm 4.7}$&      0.148$_{\pm 0.021}$&        5744$_{\pm 80}$ &       -12.74$_{\pm 0.51}$     &       0.03$_{\pm 0.02}$ & 0.08$_{\pm 0.08}$\\ 
U Sgr   &       1.605$_{\pm 0.023}$&  0.85& 30117.4812&6.745$_{\pm 6.10^{-06}}$ &1.219$_{\pm 0.028}$ &  0.757$_{\pm 0.015}$     &50.7$_{\pm 1.4}$&      0.486$_{\pm 0.021}$&        5746$_{\pm 80}$ &       2.62$_{\pm 0.50}$       &       0.00$_{\pm 0.02}$ & 0.08$_{\pm 0.08}$\\ 
BB Sgr  &       1.188$_{\pm 0.024}$&  0.82& 36053.0217&6.637$_{\pm 1.10^{-05}}$ &1.252$_{\pm 0.046}$ &  0.572$_{\pm 0.012}$     &51.8$_{\pm 1.8}$&      0.323$_{\pm 0.021}$&        5601$_{\pm 81}$ &       6.61$_{\pm 0.50}$       &       0.03$_{\pm 0.02}$ & 0.08$_{\pm 0.08}$\\ 
XX Sgr  &       0.724$_{\pm 0.027}$&  1.10& 52839.7171&6.424$_{\pm 1.10^{-05}}$ &1.375$_{\pm 0.089}$ &  0.316$_{\pm 0.007}$     &46.9$_{\pm 2.7}$&      0.633$_{\pm 0.022}$&        5952$_{\pm 81}$ &       12.39$_{\pm 0.53}$      &       0.08$_{\pm 0.02}$ & -0.01$_{\pm 0.06}$\\ 
W Sgr   &       2.402$_{\pm 0.177}$&  3.95& 48690.6794&7.595$_{\pm 1.10^{-05}}$ &1.356$_{\pm 0.107}$ &  1.180$_{\pm 0.024}$     &52.8$_{\pm 5.6}$&      0.149$_{\pm 0.021}$&        5798$_{\pm 80}$ &       -27.80$_{\pm 0.50}$     &       0.05$_{\pm 0.02}$ & 0.02$_{\pm 0.08}$\\ 
WZ Sgr  &       0.612$_{\pm 0.028}$&  0.94& 35506.5734&21.848$_{\pm 1.10^{-04}}$ &1.106$_{\pm 0.054}$ &  0.612$_{\pm 0.012}$     &107.5$_{\pm 7.3}$&     0.631$_{\pm 0.021}$&        5285$_{\pm 80}$ &       -17.25$_{\pm 0.52}$     &       0.02$_{\pm 0.02}$ & 0.28$_{\pm 0.08}$\\ 
Y Sgr   &       2.012$_{\pm 0.058}$&  1.76& 47303.1276&5.773$_{\pm 7.10^{-06}}$ &1.178$_{\pm 0.046}$ &  0.853$_{\pm 0.017}$     &45.6$_{\pm 2.1}$&      0.271$_{\pm 0.021}$&        5797$_{\pm 80}$ &       -3.03$_{\pm 0.51}$      &       0.02$_{\pm 0.02}$ & 0.05$_{\pm 0.08}$\\ 
X Sgr   &       2.843$_{\pm 0.141}$&  1.22& 48707.9150&7.013$_{\pm 1.10^{-05}}$ &1.169$_{\pm 0.062}$ &  1.345$_{\pm 0.027}$     &50.9$_{\pm 3.7}$&      0.277$_{\pm 0.021}$&        6047$_{\pm 80}$ &       -13.00$_{\pm 0.51}$     &       0.04$_{\pm 0.02}$ & -0.21$_{\pm 0.30}$\\ 
V0350 Sgr       &       0.810$_{\pm 0.062}$&  2.43& 35316.2602&5.154$_{\pm 7.10^{-06}}$ &1.783$_{\pm 0.154}$ &     0.420$_{\pm 0.008}$     &55.8$_{\pm 6.1}$&  0.337$_{\pm 0.021}$&    5803$_{\pm 80}$ &       17.72$_{\pm 0.51}$      &       0.03$_{\pm 0.02}$ & 0.18$_{\pm 0.08}$\\ 
ST Tau  &       0.916$_{\pm 0.034}$&  1.35& 41761.5441&4.034$_{\pm 3.10^{-06}}$ &1.228$_{\pm 0.074}$ &  0.303$_{\pm 0.006}$     &35.6$_{\pm 2.0}$&      0.401$_{\pm 0.021}$&        5991$_{\pm 80}$ &       1.01$_{\pm 0.51}$       &       0.08$_{\pm 0.02}$ & -0.14$_{\pm 0.15}$\\ 
RZ Vel  &       0.661$_{\pm 0.017}$&  1.24& 34845.9237&20.395$_{\pm 5.10^{-05}}$ &1.087$_{\pm 0.037}$ &  0.644$_{\pm 0.013}$     &104.7$_{\pm 4.3}$&     0.437$_{\pm 0.021}$&        5526$_{\pm 80}$ &       25.66$_{\pm 0.54}$      &       0.06$_{\pm 0.02}$ & 0.05$_{\pm 0.15}$\\ 
U Vul   &       1.308$_{\pm 0.057}$&  2.88& 48311.1041&7.991$_{\pm 2.10^{-05}}$ &1.334$_{\pm 0.067}$ &  0.777$_{\pm 0.016}$     &63.9$_{\pm 4.1}$&      0.666$_{\pm 0.021}$&        5839$_{\pm 80}$ &       -14.08$_{\pm 0.51}$     &       -0.00$_{\pm 0.02}$ & 0.19$_{\pm 0.06}$\\ 
T Vul   &       1.719$_{\pm 0.058}$&  1.20& 41704.7260&4.435$_{\pm 2.10^{-06}}$ &1.389$_{\pm 0.051}$ &  0.605$_{\pm 0.012}$     &37.9$_{\pm 2.0}$&      0.079$_{\pm 0.021}$&        5978$_{\pm 80}$ &       -1.10$_{\pm 0.50}$      &       0.06$_{\pm 0.02}$ & 0.01$_{\pm 0.08}$\\ 
S Vul   &       0.237$_{\pm 0.020}$&  1.03& 48331.9997&68.552$_{\pm 6.10^{-03}}$ &1.178$_{\pm 0.104}$ &  0.640$_{\pm 0.013}$     &290.3$_{\pm 35.6}$&    1.023$_{\pm 0.021}$&        5449$_{\pm 80}$ &       0.62$_{\pm 0.51}$       &       -0.00$_{\pm 0.02}$ & 0.12$_{\pm 0.06}$\\ 
SV Vul  &       0.402$_{\pm 0.021}$&  1.20& 48307.7580&44.941$_{\pm 1.10^{-03}}$ &1.297$_{\pm 0.070}$ &  0.835$_{\pm 0.017}$     &223.7$_{\pm 17.3}$&    0.564$_{\pm 0.021}$&        5242$_{\pm 80}$ &       -1.11$_{\pm 0.51}$      &       -0.00$_{\pm 0.02}$ & 0.05$_{\pm 0.08}$\\ 
X Vul   &       0.864$_{\pm 0.022}$&  1.06& 35308.5097&6.320$_{\pm 9.10^{-06}}$ &1.607$_{\pm 0.072}$ &  0.435$_{\pm 0.009}$     &54.1$_{\pm 2.2}$&      0.806$_{\pm 0.021}$&        5801$_{\pm 80}$ &       -14.29$_{\pm 0.51}$     &       0.04$_{\pm 0.02}$ & 0.07$_{\pm 0.08}$\\ 
\hline
\end{tabular}}
\end{table*}
\clearpage
\newpage
\section{Results of the SPIPS modeling}
\label{sec:results}
The SPIPS model fitting was performed for each of the 63 Cepheids of our sample. The final SPIPS adjustment for the Cepheid CD Cyg is presented in Fig. \ref{fig:CD_Cyg}. The available data for this Cepheid give a good example of the quality we reached for most stars of the sample. Other examples of SPIPS models are also provided in Figs. \ref{fig:SS_Sct}, \ref{fig:AQ_Pup}, \ref{fig:LS_Pup}, \ref{fig:AW_Per} \ref{fig:RS_Pup}, and \ref{fig:Delta_Cep} in the appendix for different conditions of dataset. They show the robustness of the models.

\subsection{Main parameters derived by the SPIPS algorithm}
\label{sec:plrelation}
The SPIPS algorithm returns various parameters from the modeling of each of the 63 Cepheids, such as $E(B-V)$ values, dereddened apparent and absolute magnitudes, mean radius of the star, infrared excess, and the projection factor. These parameters are provided in Table \ref{tab:parameters}.
The values agree well with those of \cite{Gallenne2021}, who derived parameters of 45 Galactic Cepheids using the SPIPS algorithm, with a fixed $p-$factor and a different dataset than the one adopted here, showing the robustness of the method. In particular, they reported that the IR excess of nearly 30\% of the Cepheids is likely produced by a CSE. We refer to this study for a detailed analysis of this effect. For our sample of stars, the uncertainties on IR excess are rather large and do not allow us to conclude about the presence of a circumstellar envelope at this stage. Adopting a precision threshold of 30\% of the IR excess value, which corresponds to the most precise values of our sample, leads to approximately the same fraction of Cepheids with detected CSEs as was observed by \cite{Gallenne2021}. In addition, given the large size of our sample, we are able to exclude a correlation between the IR excess and the period that was suggested by dusty-wind models. \cite{Hocde2020} have proposed free-free emission as an explanation for the formation of circumstellar envelopes.

The mean apparent magnitudes listed in Table \ref{tab:magnitudes} correspond to flux-averaged mean magnitudes. The $B$ and $V$ magnitudes are in the Cousins and Johnson systems, respectively. The NIR $J$, $H,$ and $K_S$ mean magnitudes, originally in the CIT system, were converted into the 2MASS system using the following transformation relations from \cite{Monson2011}, with negligible transformation errors:
\[
\begin{array}{l l l l l l l l}
K_{\rm 2MASS} = K_{\rm CIT} + 0.001_{0.005}~ (J_{\rm CIT}-K_{\rm CIT}) - 0.019_{0.004},  \\
J_{\rm 2MASS} ~= K_{\rm 2MASS} + 1.068_{0.009} ~(J_{\rm CIT}-K_{\rm CIT}) - 0.020_{0.007},   \\
H_{\rm 2MASS} = K_{\rm 2MASS} + 1.000_{0.023} ~(H_{\rm CIT}-K_{\rm CIT}) + 0.034_{0.006}.   \\
\end{array}
\]
The corresponding scatter in $J$, $H,$ and $K_s$ bands is $\sigma=0.018, 0.014,\text{and } 0.014$, respectively.
\begin{figure}[]
\centering
\includegraphics[width=8.5cm]{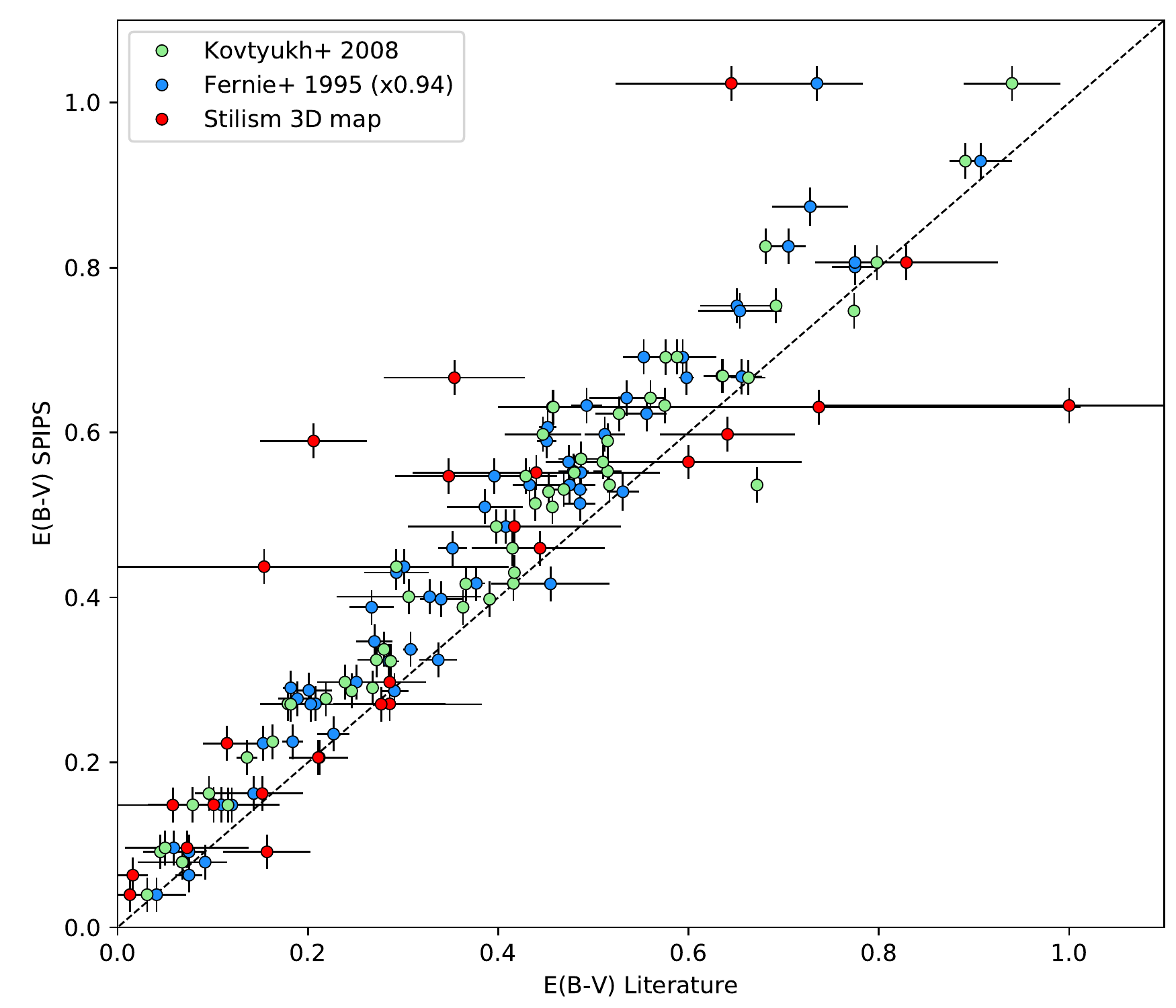}
\caption{Comparison of the color excesses derived by the SPIPS modeling with the values of the literature from \cite{Kovtyukh2008}, Stilism 3D map \citep{Lallement2018}, and \cite{Fernie1995}.}
\label{fig:EBV}
\end{figure}

    \begin{figure}[]
     \centering
     \includegraphics[width=8.5cm]{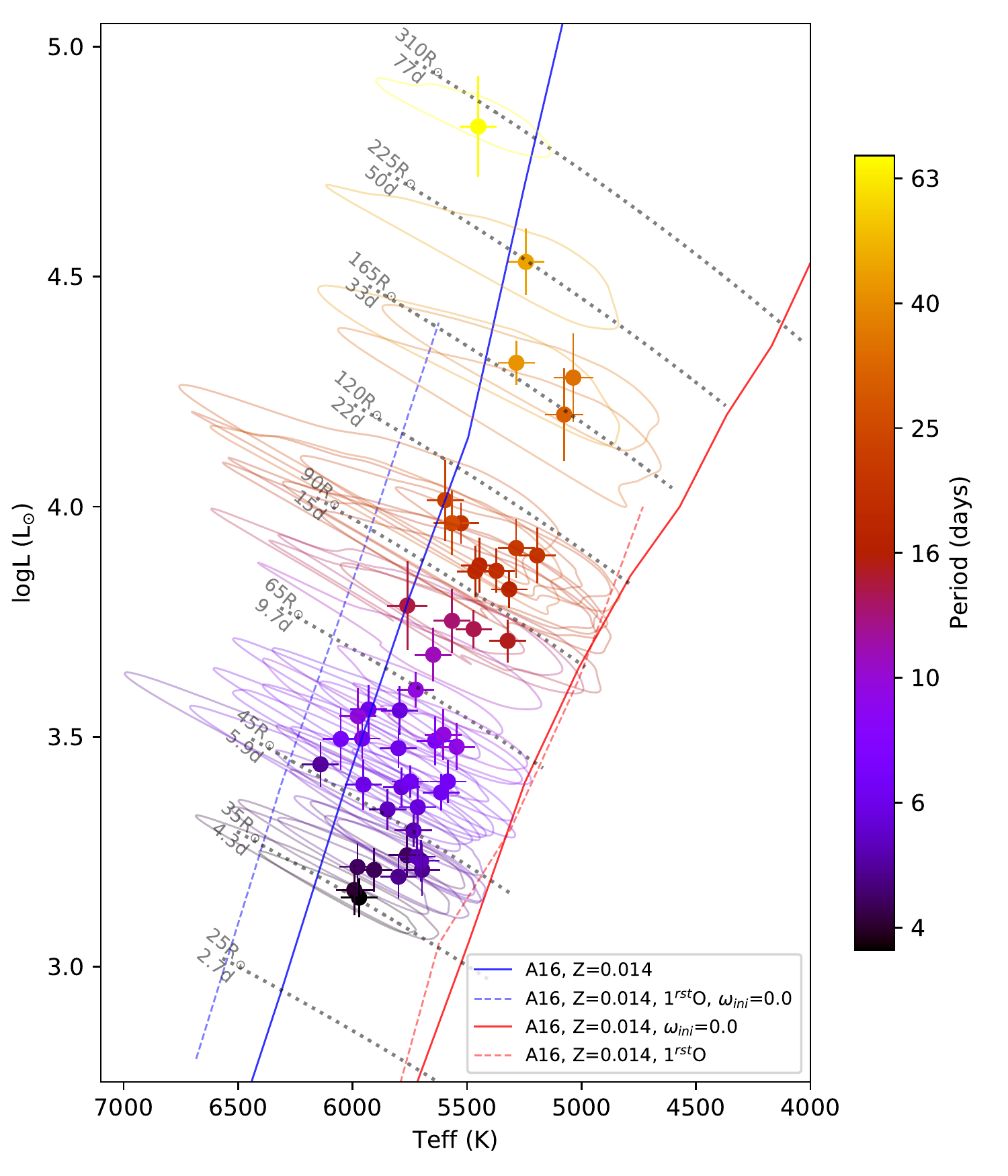}
     \caption{Position of our Galactic Cepheids (with RUWE < 1.4) during their pulsation phase in the Hertzsprung-Russell diagram. Blue and red edges are from \cite{Anderson2016a} for fundamental (solid lines) and first-overtone (dashed lines) pulsation modes and with or without rotation.}
     \label{fig:HR}
     \end{figure}

    \begin{figure*}[h!]
    \centering
    \includegraphics[width=\textwidth]{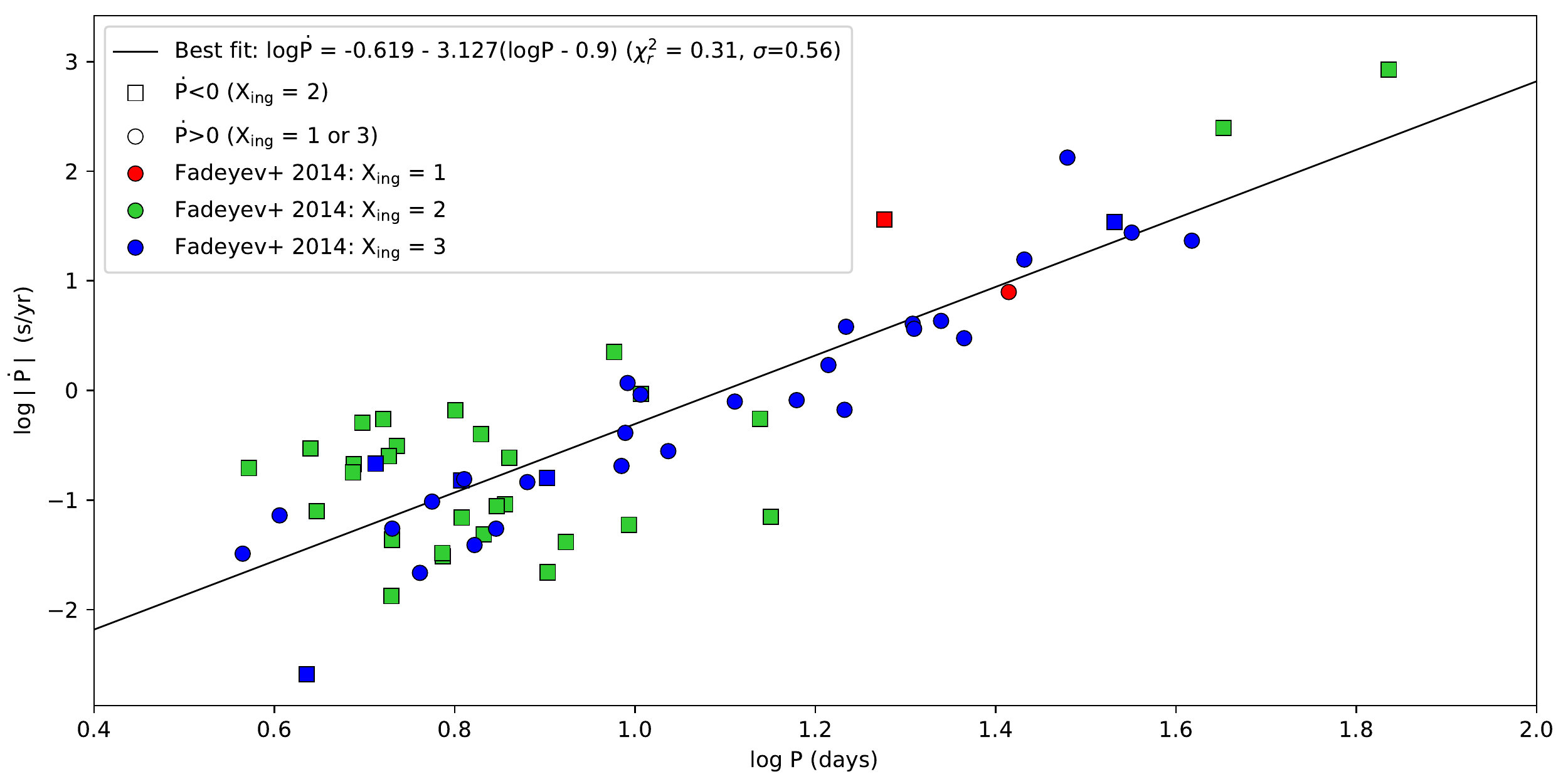}
    \caption{Dependence of the period change $\dot{P}$ on the period. X$_{\rm ing}$ indicates the crossing number as predicted by \cite{Fadeyev2014}.}
    \label{fig:sherpa}
    \end{figure*}

As mentioned in Sect. \ref{sec:spips}, the reddening $E(B-V)$ can be computed directly by the SPIPS algorithm instead of using values from the literature. Most values from literature are imprecise or derived from inhomogeneous methods, which can have a substantial effect on the consistence of the derived values and on the accuracy of the P$-$L relation calibration.
We represent in Fig. \ref{fig:EBV} a comparison of the reddening values computed by the SPIPS code with those from the Stilism 3D map \citep{Lallement2018} or derived by \citet{Kovtyukh2008} and \citet{Fernie1995}. The latter are used extensively in calibrations of the period-luminosity relation \citep{Groenewegen2018, Breuval2021}.

The dispersion between all these values is significant, but we note that the SPIPS reddening values generally agree with the others and are larger by 0.05 mag on average. The largest differences may be explained by the fact that as mentioned in Sect. \ref{sec:spips}, the reddening derived by SPIPS is based on the photometry of each star, whereas in the literature, it is obtained for a Vega-like star, which is significantly hotter than Cepheids. \cite{Hocde2020} also pointed out that ionized-gas envelopes are likely to obscure Cepheids in the optical bands from 0.05 to 0.15 mag, approximately. This difference of color excess might be due to a compensation in the SPIPS algorithm for the absence of a circumstellar envelope in the optical by an increase in $E(B-V)$. This would especially be the case for the models in which angular diameter, distance, and effective temperatures are constrained by the observations.

    \subsection{Position in the Hertzprung-Russell diagram}

    The effective temperatures and luminosities derived from our SPIPS analysis allowed us to precisely determine the variation in position of the studied Cepheids in the Hertzsprung-Russell diagram during their pulsation phase (Fig. \ref{fig:HR}). Figure 3 shows the very dynamic nature of these objects, which move significantly outside of the instability strip during a pulsation. However, the mean values of the effective temperature and the luminosity show that these objects are mainly confined between the blue and red edges defined by \citet{Anderson2016a}. Moreover, the Cepheids of our sample are closer to the blue edge on average. Although the stars at the center of the strip seem to have higher amplitudes, no strong correlation between the amplitude and the proximity of one of the edges is visible, as was reported by \cite{Fernie1990}.

    Our data cover almost 50 years of observations, which also allowed us to derive new period change rates $\dot{P}$ (listed in Table \ref{tab:periods} in the appendix). Negative period changes arise during the second crossing of the instability strip, and positive period changes correspond to a first or third crossing. These values are consistent with the predictions by \cite{Fadeyev2014}. We can note in particular that most of our Cepheids are in their second or third crossing of the instability strip. Fig. \ref{fig:sherpa} shows the linear dependence of the logarithm of the period change on the period. The scatter of this relation is mainly explained in \citet{Anderson2016a} by the rotation dependence of $\dot{P}$. \citet{Miller2020} showed that rotation is insufficient to explain this distribution of period change rates, and that other mechanisms such as mass loss are required.


    
    

    \begin{figure*}[!h]
    \centering
    \includegraphics[width=\textwidth]{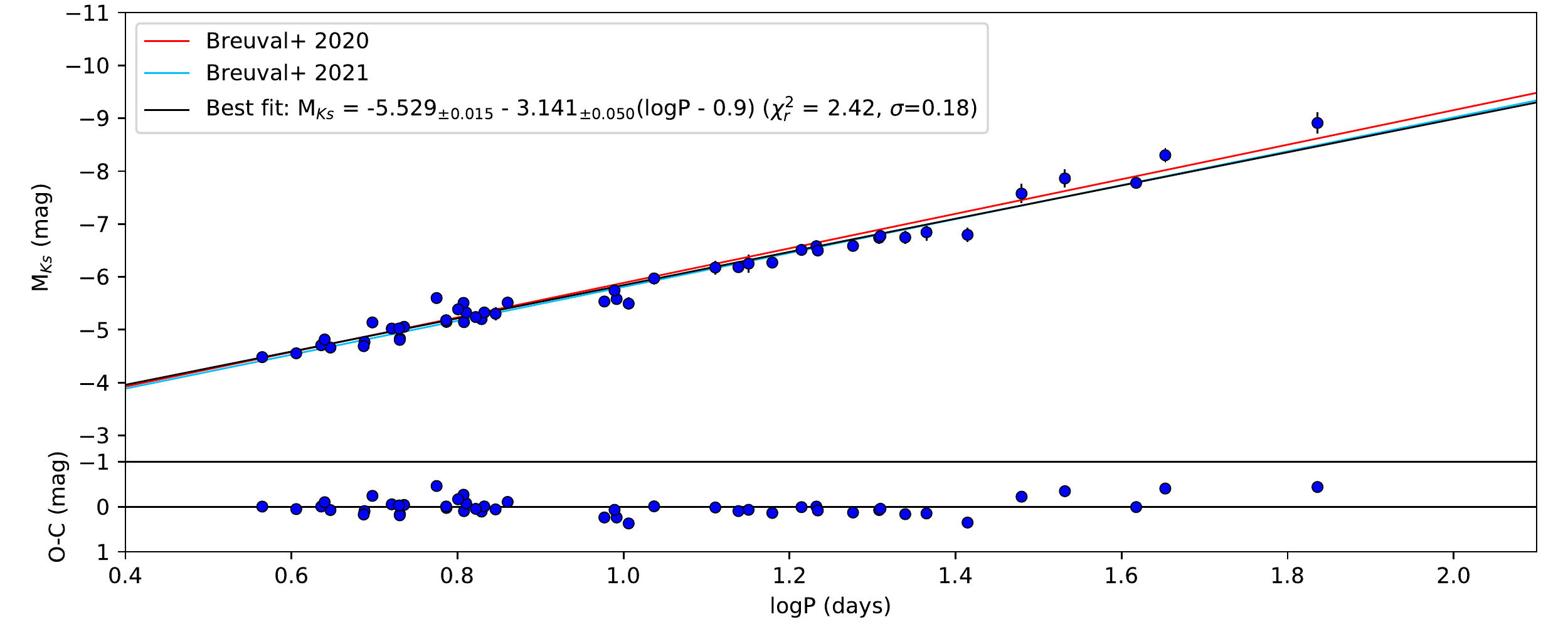}\\
    \includegraphics[width=\textwidth]{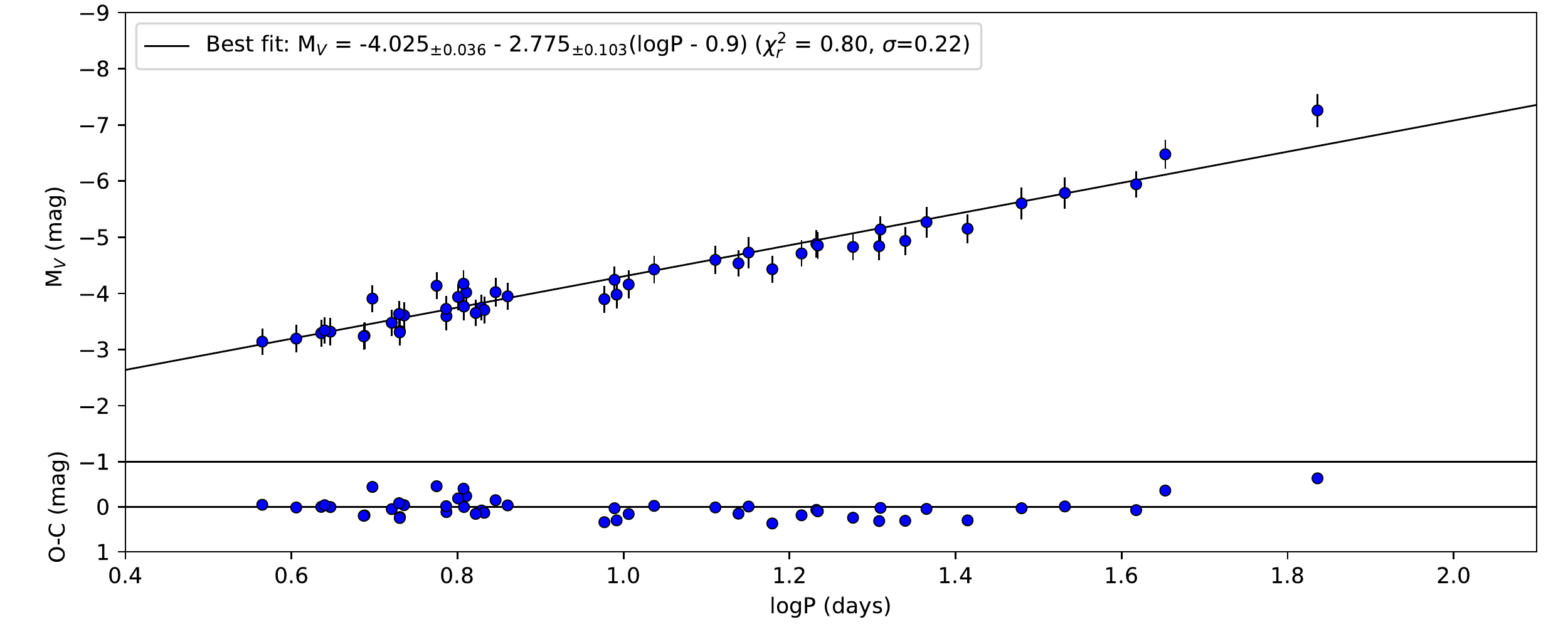}\\
    \includegraphics[width=\textwidth]{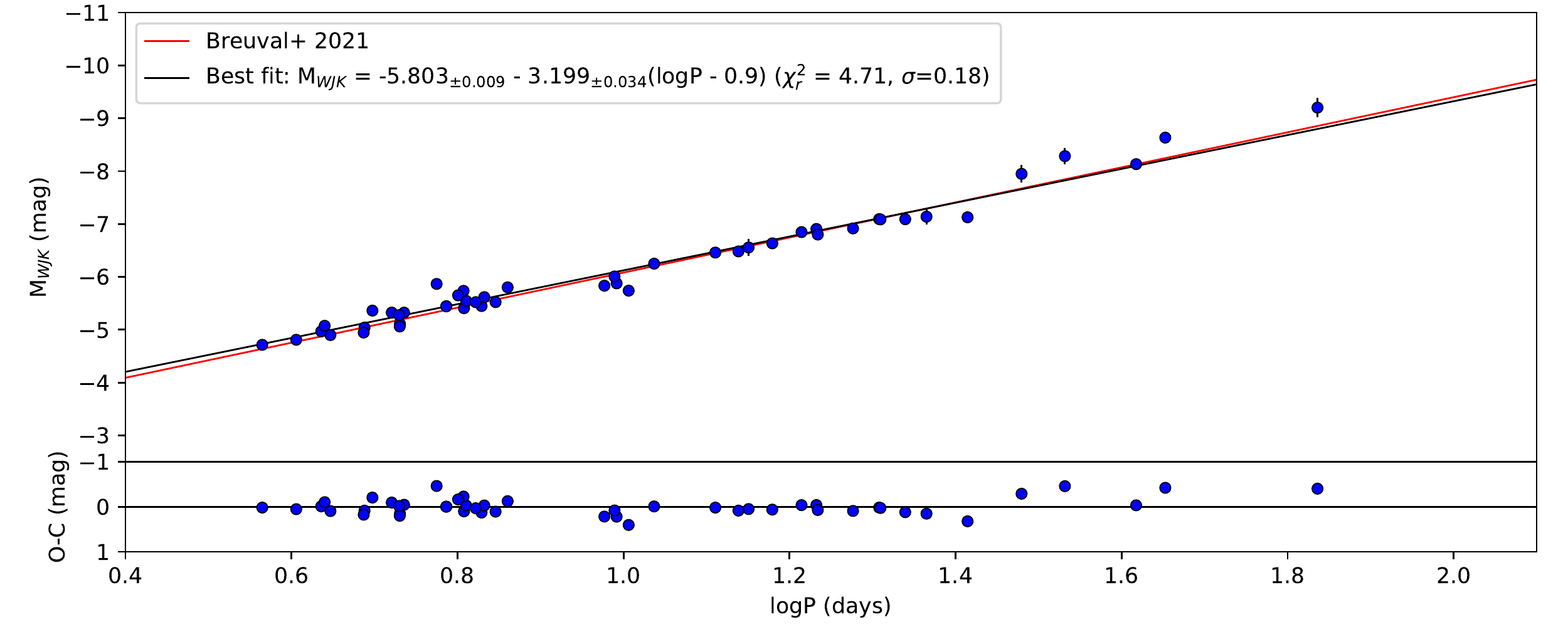}
    \caption{\footnotesize Period-luminosity relations of Galactic Cepheids in $K_S$, $V,$ and $W_{JK}$ bands ( \textbf{top}, \textbf{middle}, \textbf{and bottom}, respectively) calibrated with \textit{Gaia} EDR3 parallaxes (RUWE<1.4 only). The colored lines represent the relations from \cite{Breuval2020} and \cite{Breuval2021}.}
    \label{fig:PL}
    \end{figure*}

\begin{figure*}[]
\centering
\includegraphics[width=\textwidth]{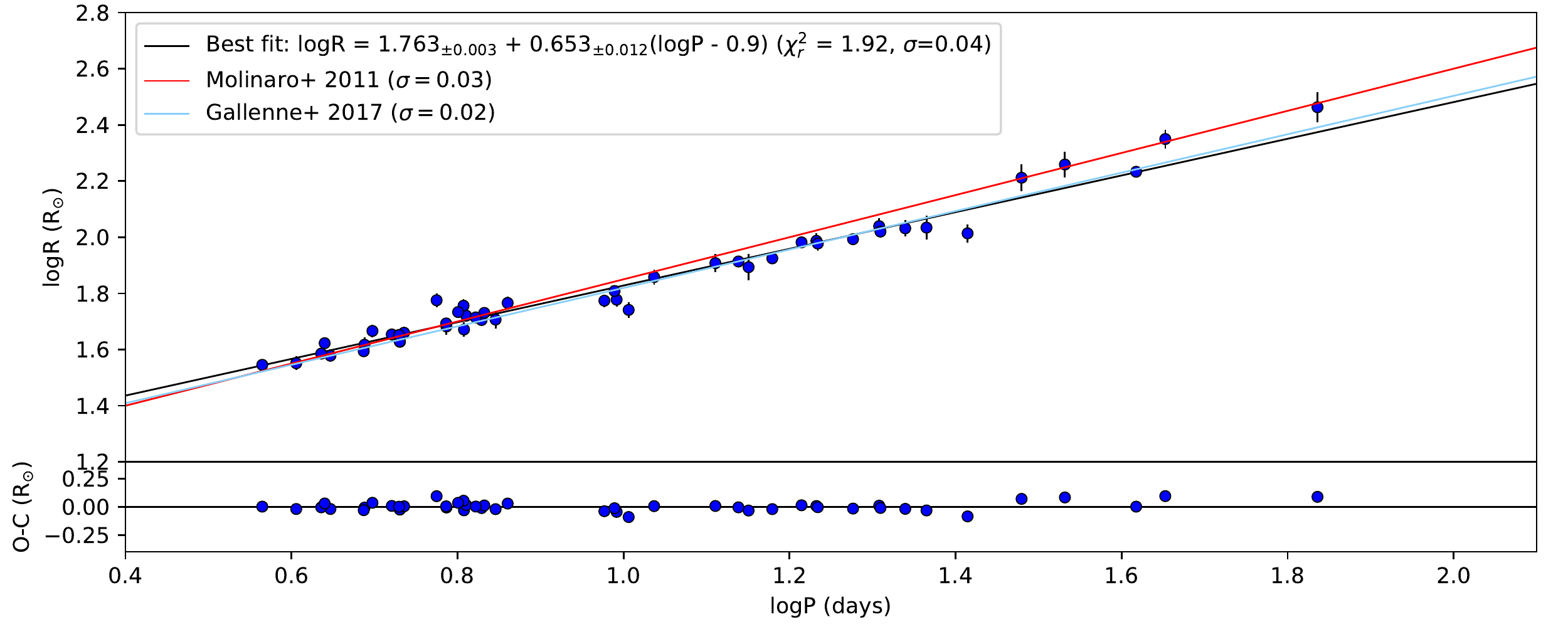}
\caption{Period-radius relation of Galactic Cepheids calibrated using \textit{Gaia} EDR3 parallaxes (RUWE<1.4 only).}
\label{fig:PR}
\end{figure*}
    
    
    

\section{Discussion}
\label{sec:discussion}

        \subsection{Period-luminosity relation from SPIPS absolute magnitudes}
        \label{sec:PL}
        
        The Cepheid period-luminosity relation is of primary importance for measuring astronomical distances. In most recent studies, this relation is the foundation of the extragalactic distance scale on which the determination of the local Hubble constant $H_0$ \citep{Breuval2020, Javanmardi2021, Riess2021} is based. Using the mean apparent magnitudes and color excesses derived by the SPIPS models, we computed the absolute magnitudes and the astrometry-based luminosities  \citep[ABL, ][]{Arenou1999} from the \textit{Gaia} EDR3 parallaxes with RUWE<1.4. 
        
        Apparent magnitudes were corrected for the extinction using the reddening law $A_{\lambda}=R_{\lambda}E(B-V)$ with $R_V = 3.10$, $R_J = 0.815$ and $R_{K_S} = 0.351$ \citep{Fitzpatrick1999}. A reddening-free Wesenheit magnitude $W_{JK}$ was also derived, defined by $W_{JK} = K_S - 0.756 (J-K_S)$.
        
        We then performed a weighted fit of the ABL function and ensure the robustness of the fit by using a Monte Carlo approach with 10000 iterations. The absolute magnitudes were parameterized around the pivot period $\log P_0=0.9$ such as $M_{\lambda} = b_{\lambda} + a_{\lambda}(\log P - 0.9)$ in order to  reduce the correlation between $a_{\lambda}$ and $b_{\lambda}$ and to minimize their respective uncertainties. We accounted for the width of the instability strip by adding in quadrature an additional term of 0.07 mag and 0.22 mag in $K_S$ and $V,$ respectively, in the magnitude errors listed in Table \ref{tab:magnitudes}.
        
        The derived PL relation in the $K_S$, $V,$ and $W_{JK}$ bands are represented in Fig. \ref{fig:PL}. The best-fit solution in the $K_S$ band corresponds to 
        \begin{equation}
        \label{eq:PL_Ks}
        M_{K_S} = -5.529_{\pm0.015}-3.141_{\pm0.050}(\log P - 0.9)
        .\end{equation}
    In $V$ and $W_{JK}$ bands, the best-fits are 
    \begin{equation}
        M_{V} = -4.025_{\pm0.036}-2.775_{\pm0.103}(\log P - 0.9)
    \end{equation}
        \begin{equation}
        W_{JK} = -5.803_{\pm0.009}-3.199_{\pm0.034}(\log P - 0.9)\text{, respectively.}
    \end{equation}
    The dispersion in the $V$, $K_S$, and $W_{JK}$ bands is 0.22, 0.18, and 0.18 mag, respectively.
        We note that our $K_S$ -band calibration agrees excellently with the result by \cite{Breuval2020} based on \textit{Gaia} DR2 parallaxes of companion stars and host open clusters and by \cite{Breuval2021} based on \textit{Gaia} EDR3 parallaxes of Cepheids (without fitting a metallicity effect).

        \subsection{Period-radius relation from SPIPS radii}
        
        In recent years, the calibration of the period-luminosity relation has been given particular importance. However, the period-radius relation of Cepheids also plays an important role in determining the masses and various parameters of these stars. As stated by \cite{Gieren1998}, this relation may also be used to derive pulsational parallaxes of Cepheids in galaxies in which radial velocity curves cannot be observed.
        For this purpose, we computed the radius of each star from the angular diameter curves modeled by the SPIPS algorithm using \textit{Gaia} EDR3 parallaxes for Cepheids with a RUWE<1.4. We derive the following period-radius relation of Galactic Cepheids, represented in Fig. \ref{fig:PR}: $$\log R = 1.763_{\pm0.003}+0.653_{\pm0.012}(\log P - 0.9).$$
        This relation has a relatively low dispersion ($\sigma=0.04$) and agrees well with the red and blue edges of the instability strip defined by \cite{Anderson2016d}. We also note that it is compatible with the relation defined by \cite{Molinaro2011} at short periods ($\log P<1$), and with the relation by \cite{Gallenne2017} established for LMC Cepheids.

        \subsection{Period-p-factor relation}

\begin{figure}[]
\centering
\includegraphics[width=8cm]{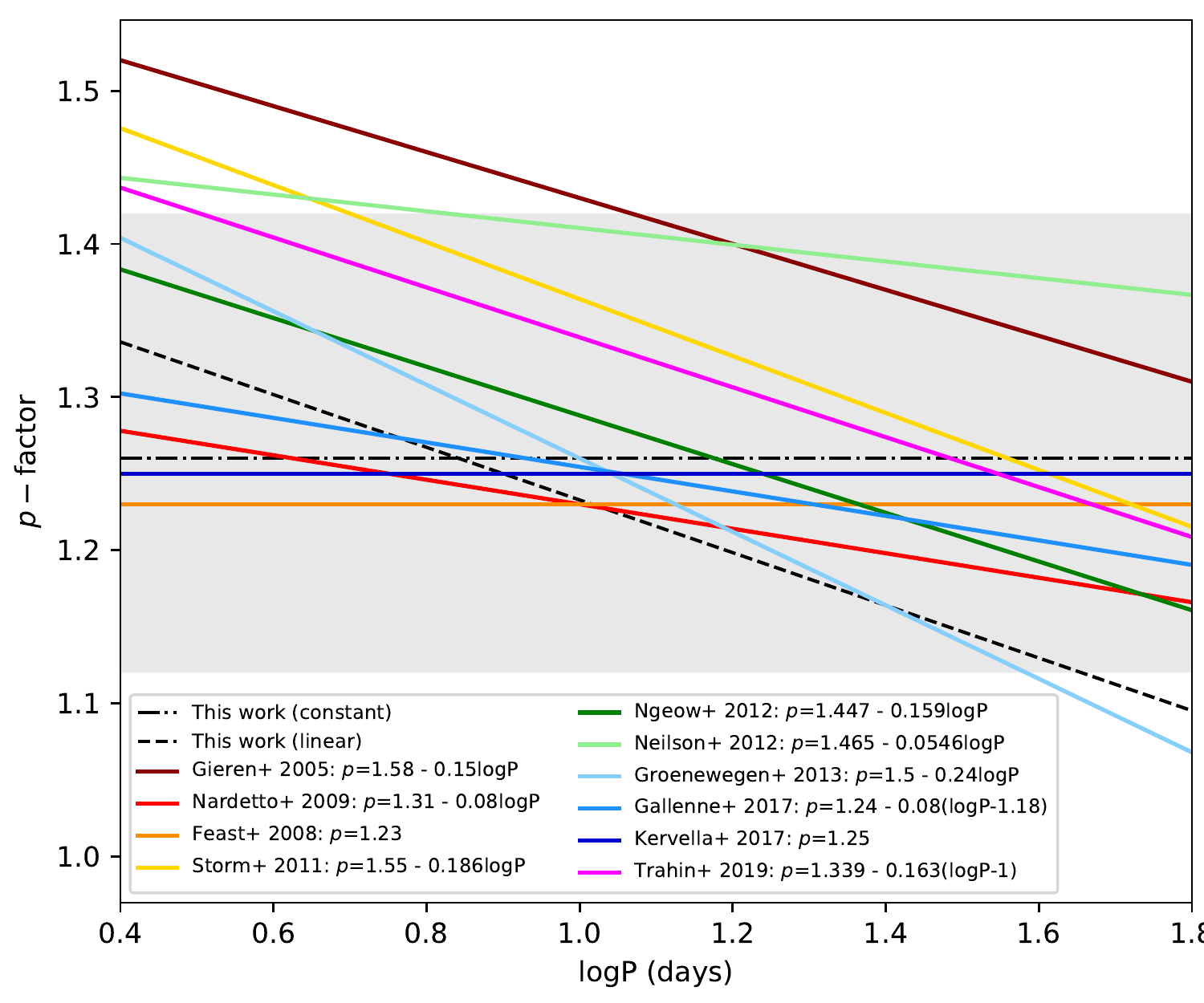}
\caption{Comparison of several period-$p-$factor relations found in the literature \citep{Gieren2005, Nardetto2009,Feast2008,Storm2011a,Ngeow2012,Neilson2012,Groenewegen2013,Gallenne2017,Kervella2017,Trahin2019}. The gray region represents the dispersion of the values around $p=1.26$ derived in this study.}
\label{fig:Pp_compare}
\end{figure}

        Many studies recently made use of the parallax-of-pulsation method with the intention of calibrating the period-luminosity relations \citep{Fouque2007, Storm2011a, Groenewegen2013, Breitfelder2016, Kervella2017, Gieren2018, Trahin2019}. As discussed in the introduction, the projection factor is still the main limitation of this method to derive accurate distances that are competitive with geometrical parallaxes. Although the physics behind this parameter is better understood nowadays through the various works by Nardetto et al. \citep{Nardetto2005, Nardetto2006, Nardetto2007, Nardetto2009, Nardetto2011, Nardetto2017}, numerous effects are still blurry. The limb-darkening is more important for the most massive stars (i.e., stars with a longer period), therefore most studies tend to conclude with a linear dependence of the $p-$factor on the period with a negative slope. Dynamical effects in the pulsating atmosphere might play a role as well. However, Fig. \ref{fig:Pp_compare} clearly shows the disparity of the P$-p$ relations found in the literature.

\begin{figure*}[]
\centering
\includegraphics[width=\textwidth]{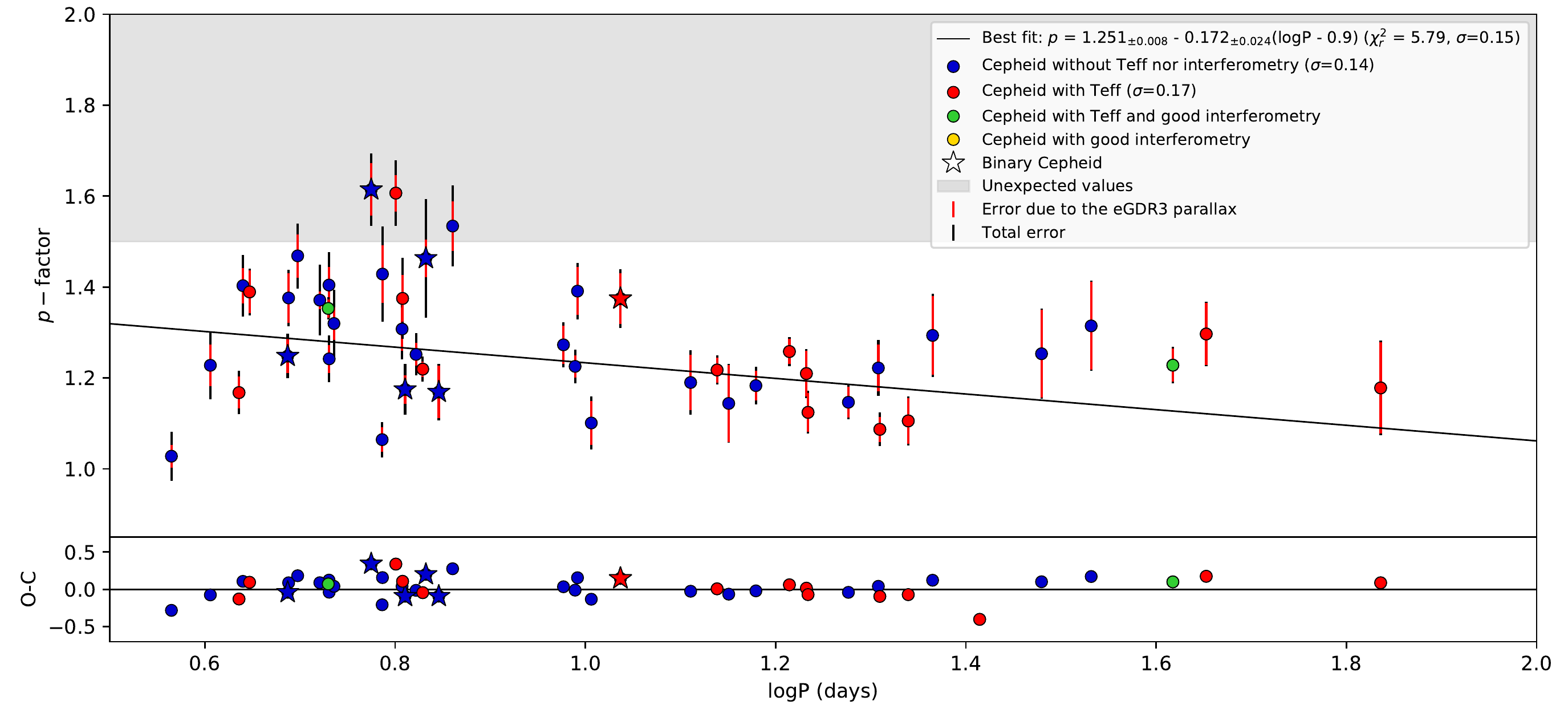}
\caption{Period$-p$-factor dependence of Galactic Cepheids using \textit{Gaia} EDR3 parallaxes (RUWE<1.4 only).}
\label{fig:Pp}
\end{figure*}

    The unprecedented precision of the recent \textit{Gaia} EDR3 parallaxes is a major tool in order to break the degeneracy of the distance over $p-$factor ratio in the PoP method and to constrain this parameter. Using the SPIPS implementation of the PoP technique described in Sect. \ref{sec:spips}, we computed the value of the projection factor for each star of our sample (with RUWE<1.4). These values are listed in the main Table \ref{tab:parameters} and are represented as a function of the logarithm of the period in Fig. \ref{fig:Pp}. We point out that the radial velocities of six Cepheids with a RUWE<1.4 are highly affected by a spectroscopic binary that can bias the results, therefore we excluded them from the fit. These stars are AW Per, VZ Cyg, V0636 Sco, X Sgr, MW Cyg, and Z Lac. Two stars ($\delta$ Cep with the \textit{Gaia} EDR3 parallax of its companion and RS Pup) have a complete dataset with a full phase coverage of interferometric angular diameters, effective temperatures, radial velocities, and multiband photometry.
    We note the high dispersion of the $p-$factor values and also some unexpected values with $p>1.5$ (area delimited in gray in Fig. \ref{fig:Pp}), which would physically correspond to a limb brightening of the stellar disk (instead of a limb darkening) or a reverse atmospheric velocity gradient (increase in velocity amplitudes toward the upper part of the atmosphere), which are highly unlikely. The uncertainties for these $p$-factors are rather large compared to best-quality $p$-factors, which suggests that the data are not optimal. On the other hand, we cannot firmly exclude any residual bias in the parallaxes, for instance, or an effect related to the CSE of Cepheids. Values lower than $p=1$ (not found in this subsample) would be physically possible if we were to consider that long-period Cepheids (and therefore Cepheids with a large radius) have stronger dynamics and an intense atmospheric velocity gradient. Finally, no dependence on the period is clearly visible, in agreement with the conclusion of the study by \cite{Pilecki2018} using Cepheids in eclipsing binaries systems. Fitting a linear relation through the points in Fig. \ref{fig:Pp} gives the following relation between the period $P$ and the $p-$factor:
    \begin{equation}
    p=1.251_{\pm 0.008}-0.172_{\pm 0.024}(\log P-0.9)
    \label{eq:Pp}
    ,\end{equation}
    with a high dispersion of 0.15. Considering only stars without effective temperatures and without interferometric measurements (blue points), we find the same dispersion of 0.14 around the same fit. Finally, stars with only an effective temperature (red points) show a scatter of 0.17 around the fit. The two stars with both effective temperature and good interferometry agree well with the slope of the fitted relation. There is no indication that one type of data is responsible for the large observed scatter. Additionally, it reinforces the robustness of the SPIPS method even for Cepheids with a limited dataset. Fitting a constant value through the points of Fig. \ref{fig:Pp} yields a projection factor of $p = 1.26 \pm 0.07$ with a dispersion of 0.15, which is not significantly higher than the dispersion obtained for Eq. \ref{eq:Pp}.
    
    Fixing this value to derive new distance estimates leads to a more dispersed PL relation: 
    \begin{equation}
    M_{Ks}=-5.488_{\pm0.037}-3.515_{\pm0.120}(\log P-0.9)
    ,\end{equation}
    with $\sigma=0.22 \, \rm mag$, which is higher by $\sim21\%$ than the previous calibration (Eq. \ref{eq:PL_Ks}). The quality criteria from \cite{Lindegren2020a} were verified for these stars, and we assume that biases due to a potential chromaticity effect \citep{Breuval2020} are negligible in the EDR3, therefore this suggests that \textit{Gaia} EDR3 parallaxes are sufficiently precise to let an intrinsic dispersion of the projection factor appear. Thus, the dispersion of the $p-$factors and the presence of values outside of the expected range suggest potential additional dependences of the P$-p$ relation, or physics of the projection factor that is still not well understood. 
    \cite{Pilecki2018} already suggested a dependence of the $p$-factor on other parameters than the period, such as the mass or radii. However, after some investigations, we did not find any correlation between the projection factor and these parameters or any other parameter, such as the mean effective temperature (Fig. \ref{fig:Teffp}), its amplitude (Fig. \ref{fig:ampTeffp}), the parallax (Fig. \ref{fig:plxp}), or the radial velocity amplitude (Fig. \ref{fig:dvradp}). Regarding the dependence on metallicity (Fig. \ref{fig:metalp}), the uncertainty of the individual values is too high to conclude about the existence of two regimes. From a theoretical point of view, \cite{Nardetto2011} predicted no correlation between the metallicity and the $p-$factor. Moreover, when we consider stars with the same period but extremely different $p-$factors in detail, no issue in the SPIPS modeling was highlighted. No correlation of radius and mass is clearly visible, in contrast to the suggestion by \cite{Pilecki2018}.
    
    A simplification made in the SPIPS algorithm is the parameterization of the infrared excess as a function of the wavelength with the assumption that there is no excess or deficit in optical bands. However, \cite{Hocde2020} showed that this effect, physically understood as being due to a circumstellar envelope, can affect not only the infrared bands, but also optical ones. 
\section{Conclusions}
\label{sec:conclusions}

We have presented the application of the SPIPS method to 63 Galactic Cepheids for which the most precise and complete dataset is available for the application of the PoP technique. This database covers almost 50 years of Cepheid observations, including multiband photometry, spectroscopic radial velocities, effective temperatures, and interferometric angular diameters.
This modeling allowed us to derive new precise and consistent mean values of several parameters such as color excesses, period changes, angular diameters, effective temperatures, multiband mean apparent magnitudes, and the $p-$factor.

We established new calibrations of the period-luminosity and period-radius relations. We finally investigated the value and dependences of the projection factor: \textit{Gaia} EDR3 parallaxes did not allow us to highlight a significant correlation between the $p-$factor and the period, but rather indicated that the $p-$factor is consistent with a constant value of $p = 1.27 \pm 0.06$, with a significant dispersion of 0.15. This dispersion and the presence of unexpected $p-$factor values suggest that other important physical phenomena affect the PoP technique that have not yet been identified. Additionally, this study suggests that the period-$p-$factor relation may have an intrinsic width and/or may depend on many individual properties. However, its physical origin is still unknown and should be investigated in the future. We found no correlations between the $p$-factor or other parameters such as the mass, radius, effective temperature, or metallicity.

There are still several aspects to overcome before the $p-$factor is understood, and the very first is probably to wait for the final \textit{Gaia} data release to obtain the best parallaxes possible in terms of precision and accuracy. In particular, improved \textit{Gaia} distances in the next releases for Cepheids with many interferometric observations such as $\delta$ Cep, RS Pup, $\beta$ Dor, or $\zeta$ Gem would permit us to obtain a better constraint on the $p-$factor. Another aspect to improve is the measurement of atmospheric velocity gradient using dedicated contribution functions of the line-forming regions. One of the best hopes is also related to the environment of Cepheids: recent studies appear to show that the circumstellar environment of Cepheids might not be static and may have some effects in the optical domain, and most probably in a different way, depending on the position of the Cepheids in the instability strip. This might explain the dispersion that we observe in the p-factors. This means that before we model the Cepheids and the $p-$factor in greater detail, we first need to understand the general scheme of the physics of the close circumstellar environnements of Cepheid through ongoing Cepheid observations in the NIR (MATISSE/VLTI) and optical (CHARA/SPICA) domains.
Moreover, parallel independent applications of the PoP technique would allow us to understand the physics of pulsating stars in more detail in order to conclude about the reliability of this method for the calibration of the extragalactic distance scale.\\

\begin{acknowledgements}
We thank Bogumil Pilecki for his careful reading that helped improved the present paper. The research leading to these results has received funding from the European Research Council (ERC) under the European Union’s Horizon 2020 research and innovation programme under grant agreement No 695099 (project CepBin). The authors also acknowledge the support of the French Agence Nationale de la Recherche (ANR), under grant ANR-15-CE31- 0012-01 (project UnlockCepheids).
This work has made use of data from the European Space Agency (ESA) mission \textit{Gaia}, processed by the \textit{Gaia} Data Processing and Analysis Consortium (DPAC). Funding for the DPAC has been provided by national institutions, in particular the institutions participating in the \textit{Gaia} Multilateral Agreement. This research made use of Astropy7, a community-developed core Python package for Astronomy \citep{Astropy-Collaboration2018}. We also used the SIMBAD and VIZIER databases and catalog access tool at the CDS, Strasbourg (France), and NASA’s Astrophysics Data System Bibliographic Services. We also aknowledge the SVO Filter Profile Service (http://svo2.cab.inta-csic.es/theory/fps/) supported from the Spanish MINECO through grant AYA2017-84089. We finally acknowledge with thanks the variable star observations from the AAVSO International Database contributed by observers worldwide and used in this research.
\end{acknowledgements}

\bibliographystyle{aa}
\bibliography{SPIPS_EDR3.bib}

\begin{appendix} 
\clearpage

\section{References of the complete dataset}
\onecolumn

\setlength{\tabcolsep}{8pt}
\begin{longtable}{l l l l l}
\caption{References of the data available for the sample of MW Cepheids.}\\
\hline
\hline
Star & Photometry & Radial Velocity & T$_{\textrm{eff}}$ & Ang. Diam. \\ 
\hline
\endfirsthead
\multicolumn{5}{c}{\textbf{Table \ref{tab:refs}} (continued)} \\
\hline
\hline
Name & Photometry & Radial Velocity & T$_{\textrm{eff}}$ & Ang. Diam. \\ 
\hline 
\endhead
\hline
\endfoot
\endlastfoot
V1162 Aql &1,2,3,4,5    &6,7    &8      &-\\ 
TT Aql &1,2,3,4,5,9,10,11,12,13,14,15   &7,16,17,18,19  &8,20   &21,22\\ 
U Aql$^{\star\bullet}$ &1,2,3,5,11,12,13,14,15  &17     &8      &21,23\\ 
FM Aql &1,2,3,4,5,9,10,11,13,15 &7,17,19        &8      &-\\ 
SZ Aql &1,2,3,4,5,9,10,11,13,14,15,24   &19     &8,20   &22\\ 
FN Aql &1,2,3,4,5,10,11,13      &7,19   &8      &-\\ 
$\eta$ Aql$^{\bullet}$ &1,3,9,10,11,12,13,14,15 &7,12,16,17,19  &8,25,26        &21,27,28,29\\ 
SY Aur &1,2,3,4,5,9,11,30       &31     &8      &-\\ 
RT Aur$^{\bullet}$ &1,2,3,4,5,9,10,11,12,14,32  &7,12,17        &8,33   &22,23\\
VY Car &1,4,5,13,15,24,30,32,34 &31     &8      &-\\ 
$\ell$ Car$^{\bullet}$ &1,3,5,14,15,24,34       &17,35  &8      &21,29\\ 
DD Cas &1,2,3,4,5,9,11  &7,31,36        &8      &-\\ 
CF Cas &1,2,3,4,5,11,14,30      &7,36,37,38     &8      &-\\ 
SW Cas &1,2,3,4,5,9,11,32       &7      &8      &-\\ 
DL Cas$^{\star\bullet}$ &1,2,3,4,5,9,11,14,30   &7,17,36,37     &8      &-\\ 
KN Cen &1,3,4,5,15,24,30,34,39  &31,38  &       -&-\\ 
$\delta$ Cep$^{\star\bullet}$ &1,3,5,9,10,11,14,40      &7,16,17,19,36  &8,33   &27\\ 
V0459 Cyg &1,2,3,4,5,11 &7,37,38        &8      &-\\ 
SZ Cyg &1,2,3,4,5,9,11,30       &31     &8      &-\\ 
V0538 Cyg &1,2,3,4,5,9  &7      &8      &-\\ 
V0402 Cyg &1,2,3,4,5,9,11       &7,37   &8      &-\\ 
CD Cyg &1,2,3,4,5,9,11,13,14,30 &7,17,31        &8,20   &-\\ 
X Cyg &1,3,4,5,9,10,11,12,13,14,30      &7,12,16,17,19,36       &8,20   &22\\ 
MW Cyg$^{\star}$ &1,2,3,4,5,9,11        &7      &8      &-\\ 
V0386 Cyg &1,2,3,4,5,9,11       &7,37   &8      &-\\ 
VZ Cyg$^{\star}$ &1,2,3,4,5,9,10,11,13  &16,17,19       &8      &-\\ 
$\beta$ Dor$^{\bullet}$ &1,3,5,14,15,24 &17,18  &8,26   &21,29,41\\ 
$\zeta$ Gem$^{\bullet}$ &1,3,5,9,11,12,14,42    &7,12,16,17,36  &8,25,26        &28\\ 
V Lac &1,2,3,4,5,9,11   &7      &8      &-\\ 
BG Lac$^{\bullet}$ &1,2,3,4,5,10,11     &19     &8      &-\\ 
RR Lac &1,2,3,4,5,9,11,30,32    &7,36   &8      &-\\ 
Z Lac$^{\star}$ &1,2,3,4,5,9,10,11,30,32        &7,19   &8,20   &-\\ 
Y Lac &1,2,3,4,5,9,10,11        &17,19  &8,33   &-\\ 
CV Mon &1,2,3,4,5,9,11,14,15,24 &7,16,37        &8      &-\\ 
T Mon$^{\star\bullet}$ &1,2,3,4,5,9,11,13,14,15,24,30,43        &7,16,17,18,36  &8,20   &21,22\\ 
S Mus$^{\star\bullet}$ &1,3,4,5,13,24,32,39     &17,18  &8      &21\\ 
S Nor &1,3,4,5,13,14,24,30,39   &17,36  &8      &21\\ 
AW Per$^{\star}$ &1,2,3,4,5,9,11,12     &7,17,44        &8      &23\\ 
RS Pup &1,3,4,5,11,13,15,24,30,34,45    &17,35  &8,26   &21\\ 
AQ Pup &1,3,4,5,11,24,30,32,34,46       &16,31  &       -&-\\ 
VZ Pup &1,3,4,5,15,24,30,32,34,43,46    &16,17  &8      &-\\ 
X Pup &1,3,4,5,11,15,24,30,34   &16,18,31       &8,20   &-\\ 
LS Pup &1,3,4,5,24,46   &6,16   &       -&-\\ 
RY Sco &1,4,5,11,13,15,24,30,34 &31     &8      &-\\ 
V0636 Sco$^{\star}$ &1,3,4,5,13,32,39   &17,18  &8      &-\\ 
SS Sct &2,3,4,5,11,13,15        &6,47   &8      &-\\ 
Z Sct &2,3,4,5,11,15,30,34      &31     &8      &-\\ 
S Sge$^{\star\bullet}$ &1,3,4,5,9,10,11,12,13,14,39     &7,12,17,19     &8,25   &21,22\\
U Sgr &1,2,3,4,5,11,13,14,15,24,30,32   &7,16,17,36     &8,25   &21\\ 
BB Sgr &1,3,4,5,11,13,15,24,32  &7,16   &8      &-\\ 
XX Sgr &1,3,4,5,11,13   &6,16   &8,26   &-\\ 
W Sgr$^{\star\bullet}$ &1,3,4,5,11,13,14,32,39  &17,18,36       &8,25   &-\\ 
WZ Sgr &1,2,3,4,5,11,13,14,24,30,34,39,43       &7,31   &8,20   &-\\ 
Y Sgr$^{\bullet}$ &1,3,4,5,11,13,14,15,32       &6,16,17        &8,26,33        &21\\ 
X Sgr$^{\star}$ &1,3,4,5,11,13,14,15,32,42      &16,17  &8      &-\\ 
V0350 Sgr$^{\star\bullet}$ &1,3,4,5,11,13       &6,7,17,18      &8      &21\\ 
ST Tau &1,2,3,4,5,10,11,32      &7,36   &8      &-\\ 
RZ Vel &1,3,4,5,15,24,32,34     &17,48  &8,26   &21\\ 
U Vul$^{\star\bullet}$ &1,2,3,4,5,9,10,11,12,14 &7,17,19,36     &8,25   &21,22\\
T Vul &1,3,4,5,9,10,11,12,13,14,32      &12,17,19,36    &8,33   &-\\ 
S Vul &1,2,5,13,14,24,30        &7,17,31        &8,20   &21,22\\ 
SV Vul &1,2,3,4,5,9,10,11,12,13,14,24,30        &7,16,17,19,36  &8,20   &21,22\\
X Vul &1,2,3,4,5,9,11   &7,36   &8,25   &-\\
\hline
\label{tab:refs}
\end{longtable} 
\tablefoot{A star$^{\text{}}$ indicates Cepheids whose radial velocities are affected by a spectroscopic companion (see Sect. \ref{sec:data}). A bullet point$^{}$ indicates stars with RUWE>1.4.}
\tablebib{(1)\cite{Berdnikov2008}; (2)\cite{Monson2011}; (3)\cite{ESA1997b}; (4)\cite{Gaia-Collaboration2018a}; (5)\cite{Cutri2003}; (6)\cite{Groenewegen2013}; (7)\cite{Gorynya1995}-\cite{Gorynya1998}; (8)\cite{Luck2018}; (9)\cite{Szabados1977}-\cite{Szabados1991}; (10)\cite{Barnes1997}; (11)\cite{Moffett1984}; (12)\cite{Kiss1998}; (13)\cite{Welch1984}; (14)\cite{Monson2012}; (15)\cite{Pel1976}; (16)\cite{Storm2011a}; (17)\cite{Borgniet2019}; (18)\cite{Petterson2005}; (19)\cite{Barnes2005}; (20)\cite{Kovtyukh2005}; (21)PIONIER; (22)CLASSIC; (23)MIRC; (24)\cite{Laney1992}; (25)\cite{Luck2004}; (26)\cite{Proxauf2018}; (27)FLUOR; (28)PTI; (29)VINCI; (30)\cite{Harris1980}; (31)\cite{Anderson2016}; (32)AAVSO; (33)\cite{Andrievsky2005}; (34)\cite{Madore1975}; (35)\cite{Anderson2014}; (36)\cite{Bersier1994}; (37)\cite{Metzger1991,Metzger1992}; (38)\cite{Pont1994}-\cite{Pont1996}; (39)\cite{Walraven1964}; (40)\cite{Engle2014}; (41)SUSI; (42)\cite{Feast2008}; (43)\cite{Coulson1985}; (44)\cite{Evans2000}; (45)\cite{Kervella2017}; (46)\cite{Schechter1992}; (47)\cite{Groenewegen2008}; (48)\cite{Szabados2015}.}
\clearpage
\newpage

\section{Examples of SPIPS fits}

\begin{figure*}[h!]
\centering
\includegraphics[width=17cm]{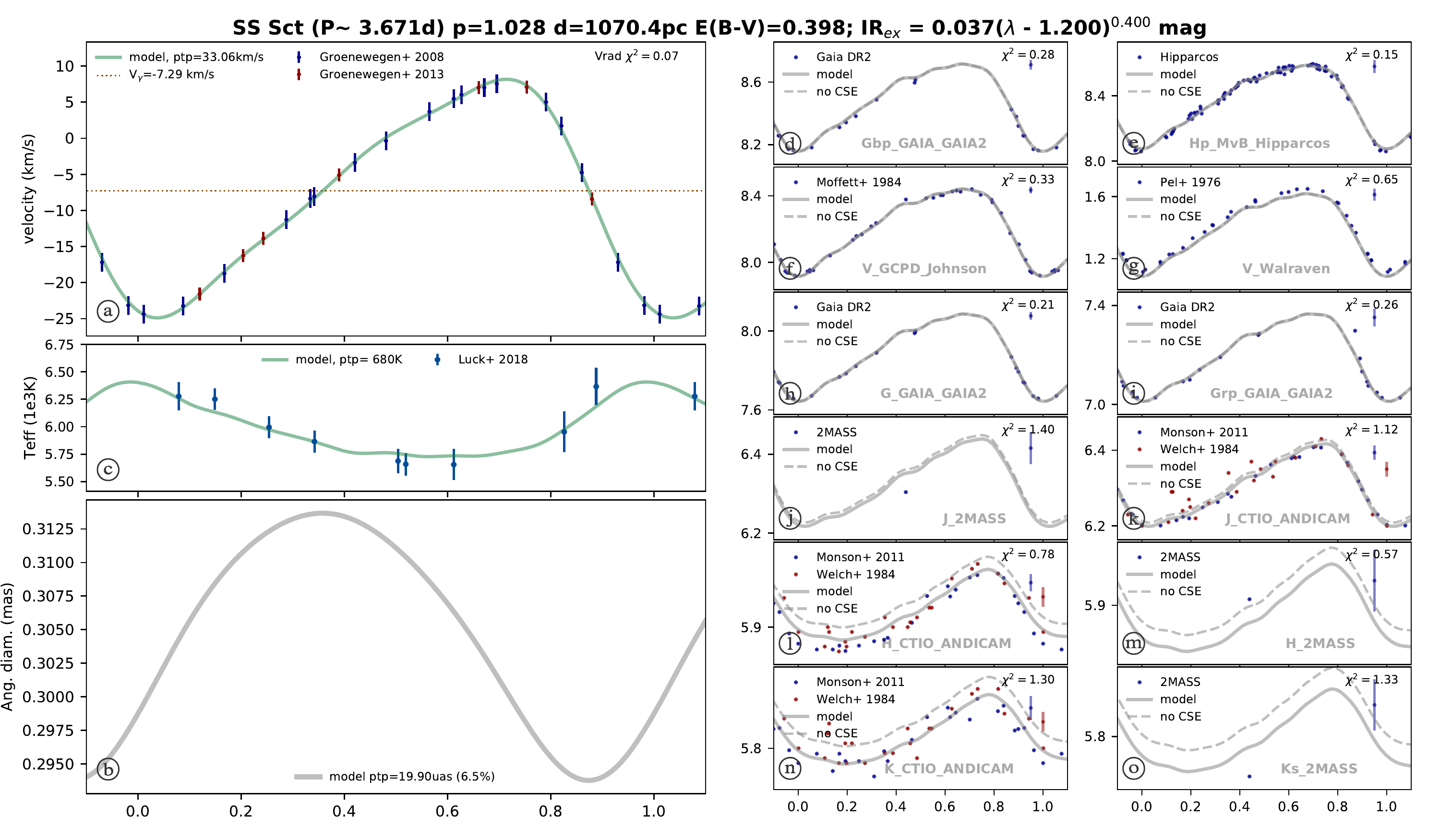}
\caption{Result of the SPIPS modeling for the Cepheid SS Sct (P=3.67 days). This star is representative of a bad dataset in the adopted sample that has only a few radial velocity data and dispersed photometry.}
\label{fig:SS_Sct}
\end{figure*}

\begin{figure*}[h!]
\centering
\includegraphics[width=17cm]{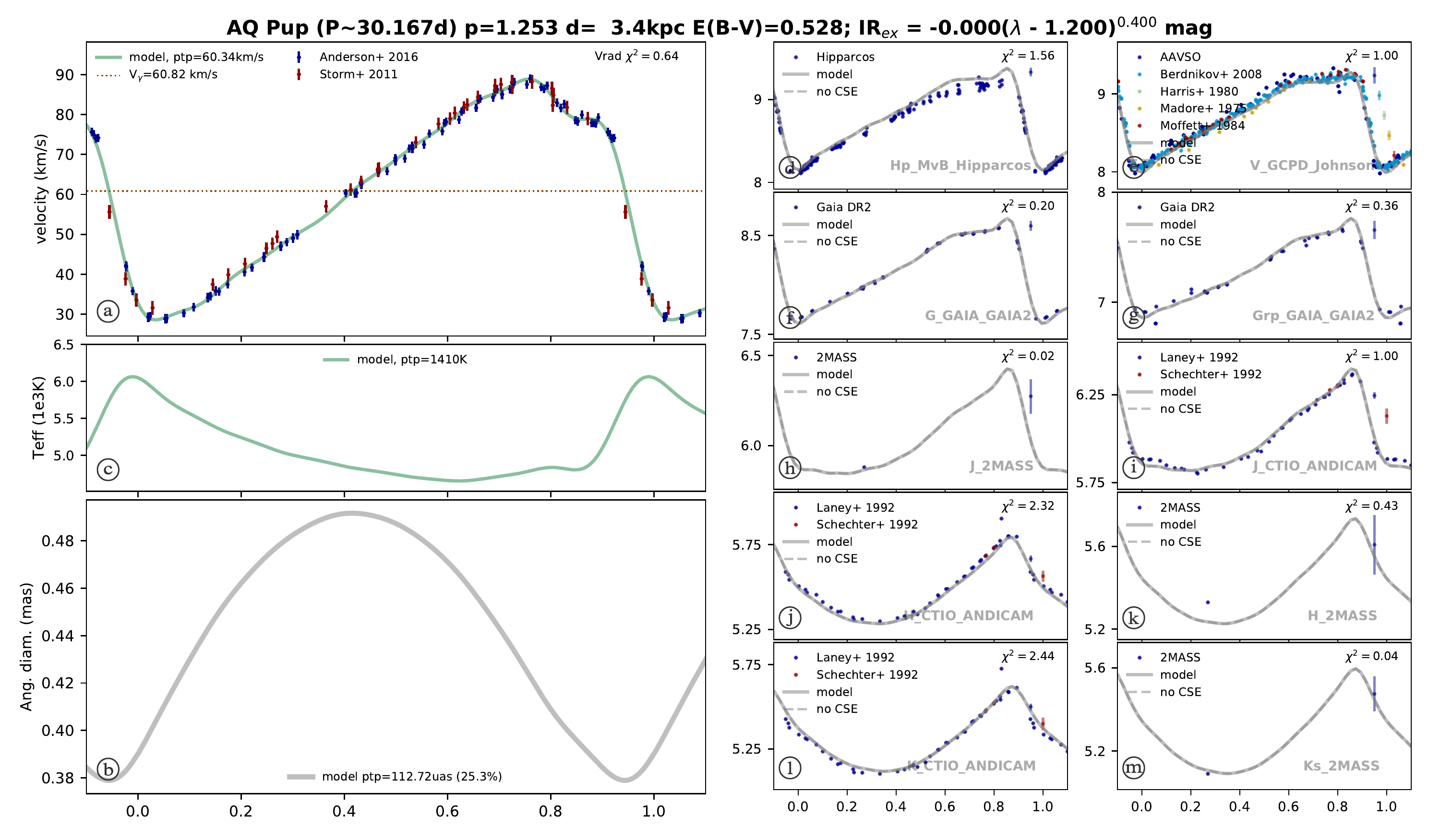}
\caption{Result of the SPIPS modeling for the Cepheid AQ Pup (P=30.17 days). Only multiband photometry and radial velocities are available for this star, but with a full phase coverage and a low dispersion.}
\label{fig:AQ_Pup}
\end{figure*}

\clearpage
\newpage

\begin{figure*}[h!]
\centering
\includegraphics[width=17cm]{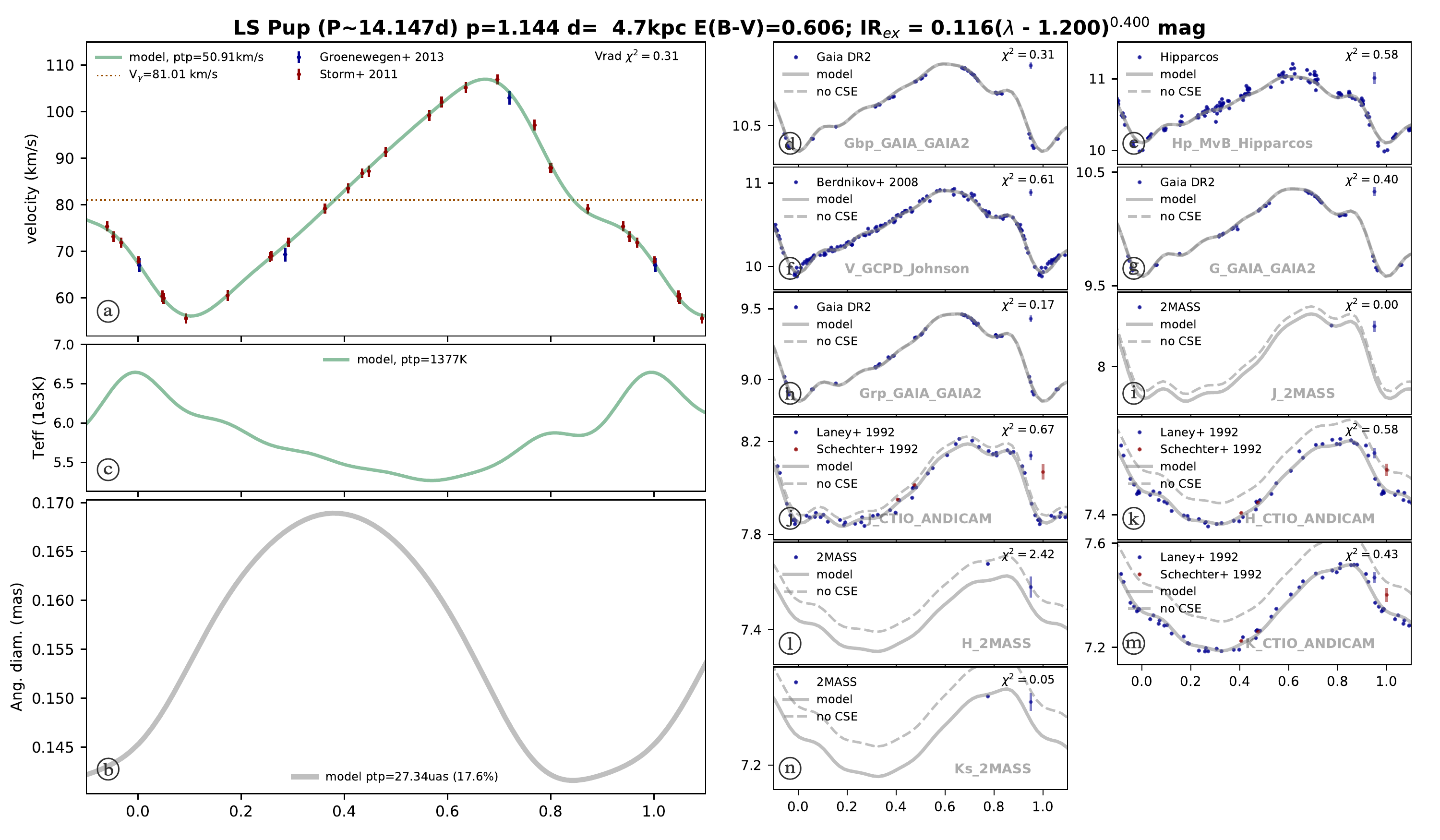}
\caption{Result of the SPIPS modeling for the Cepheid LS Pup (P=14.15 days). Only multiband photometry and radial velocities are available for this star, but with a full phase coverage and a low dispersion. Moreover, its light curves present a strong bump, which makes the adjustment more complex.}
\label{fig:LS_Pup}
\end{figure*}

\begin{figure*}[h!]
\centering
\includegraphics[width=17cm]{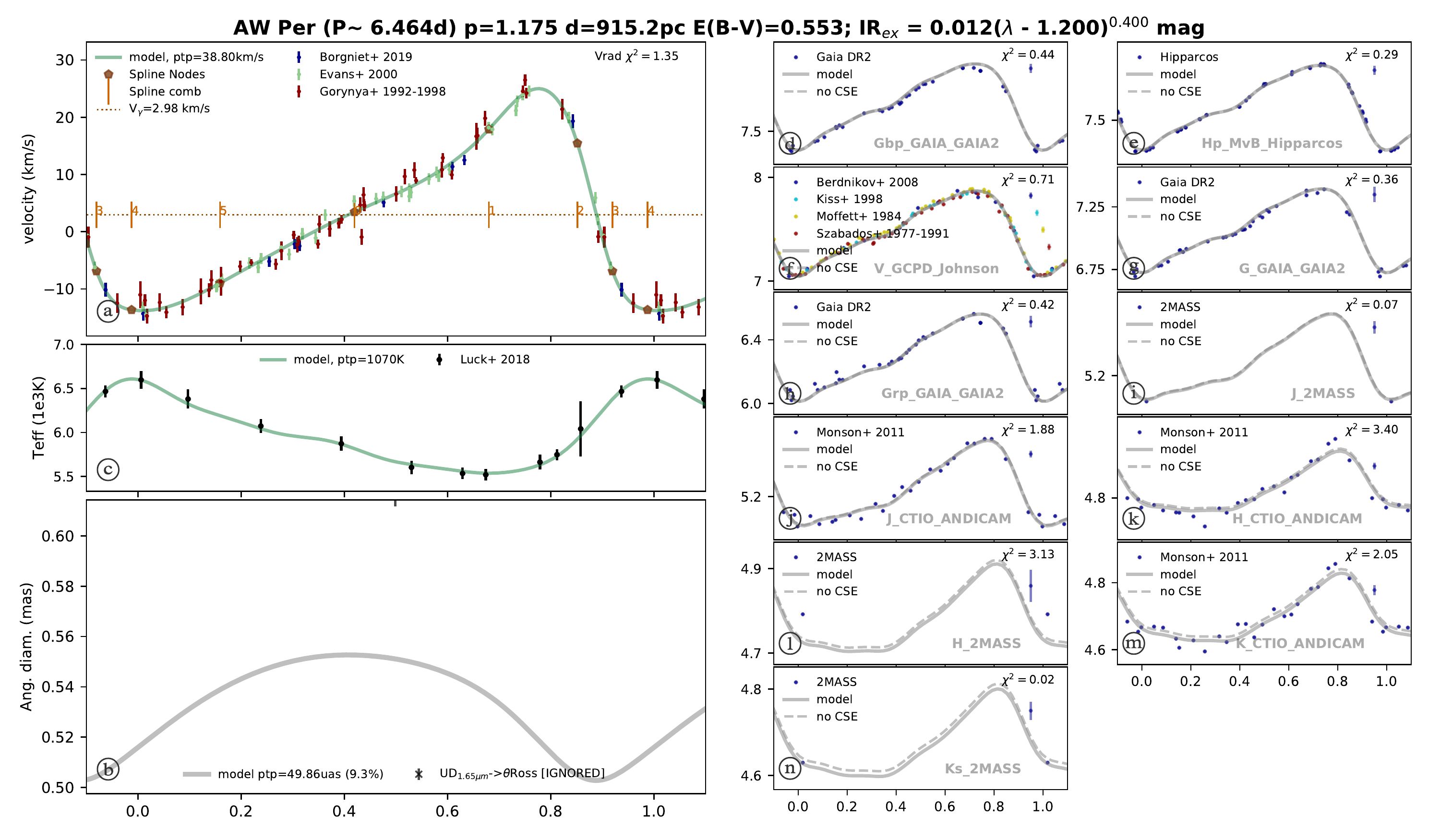}
\caption{Result of the SPIPS modeling for the Cepheid AW Per (P=6.46 days). Spectroscopic effective temperatures and full phase coverage multiband photometry and radial velocities are available. As for CD Cyg represented in the main body of the present paper, the dataset of this star is representative of the quality we reached for most stars of the sample.}
\label{fig:AW_Per}
\end{figure*}

\clearpage
\newpage

\begin{figure*}[h!]
\centering
\includegraphics[width=17cm]{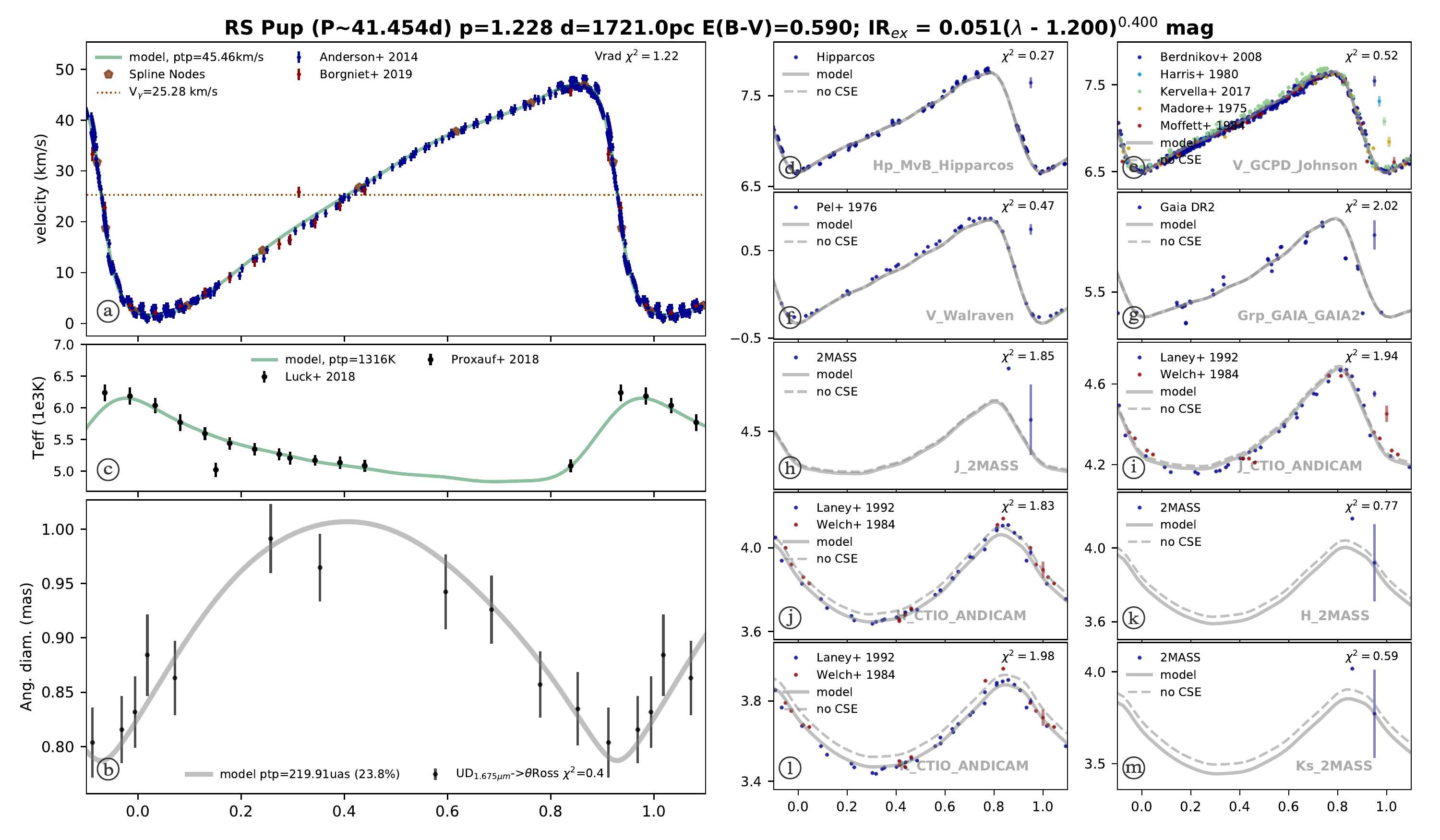}
\caption{Result of the SPIPS modeling for the Cepheid RS Pup (P=41.45 days). Through a complete and precise dataset associated with an accurate EDR3 parallax, this star represents one of the best adjustments available in this sample. }
\label{fig:RS_Pup}
\end{figure*}

\begin{figure*}[h!]
\centering
\includegraphics[width=17cm]{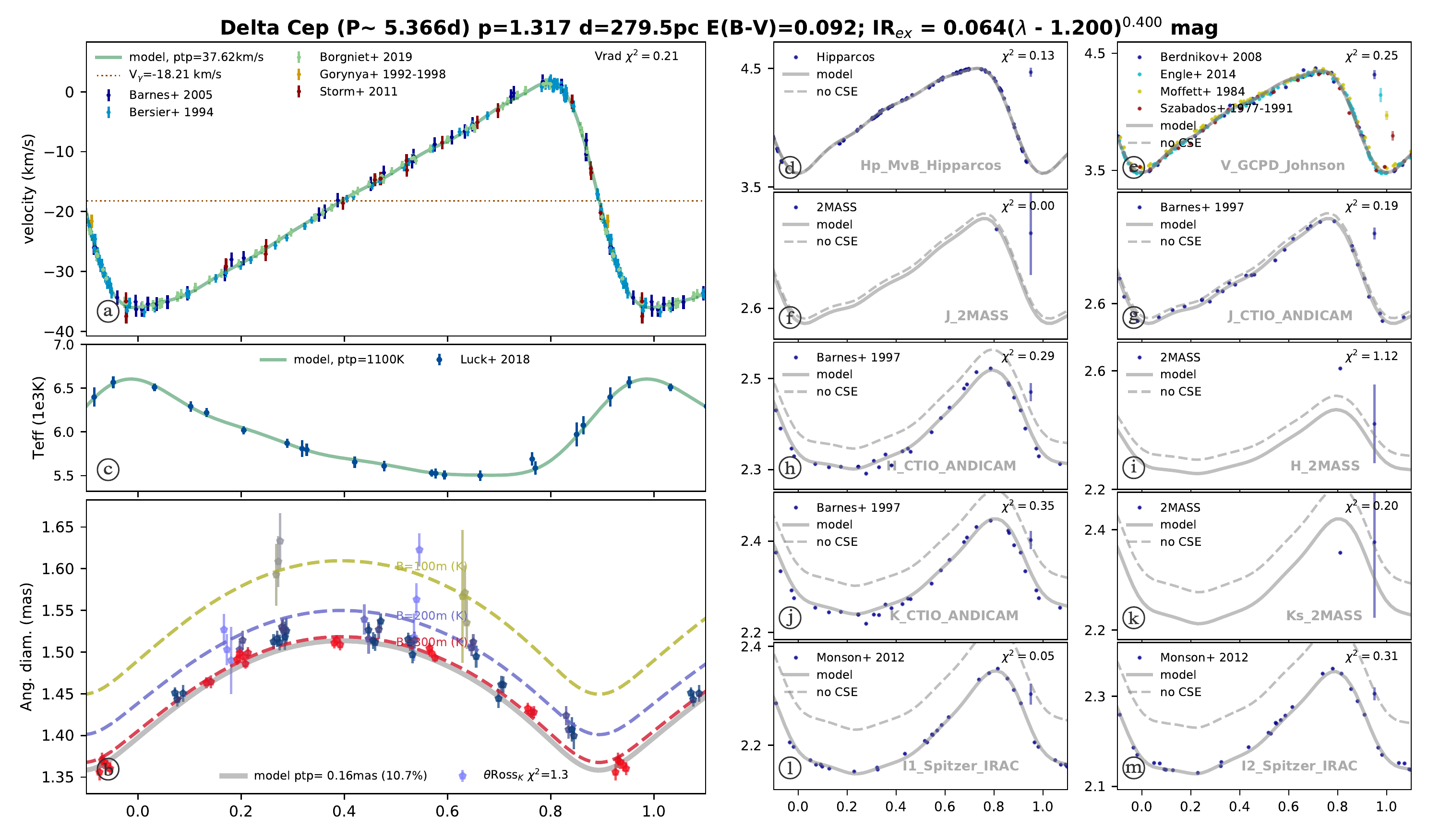}
\caption{Result of the SPIPS modeling for the Cepheid $\delta$ Cep (P=5.36 days). This star has the most complete dataset available, with interferometric angular diameters, spectroscopic effective temperatures, and full phase coverage multiband photometry and radial velocities from many studies. The \textit{Gaia} EDR3  parallax used in this adjustment is unreliable, with a RUWE parameter of 2.71. In order to take advantage of its data, we present in Table \ref{tab:parameters} the results of the adjustment using the accurate \textit{Gaia} EDR3 parallax of its companion derived by \citep{Kervella2019b}.}
\label{fig:Delta_Cep}
\end{figure*}

\clearpage
\newpage

\section{Mean apparent magnitudes derived from SPIPS modeling}

\setlength{\tabcolsep}{6pt}
\begin{longtable}{l c c c c c c c c c c c c c c c c }
\caption{Flux-averaged mean apparent magnitudes (not dereddened) derived from the SPIPS modeling in visible and infrared bands.}\\
\hline
\hline
Star & $B$ & $G_{\rm BP}$ & $V$ & $G$ & $G_{\rm RP}$ & $J$ & $H$ & $K_{S}$ & $I_1$ & $I_2$\\
 & (mag) & (mag) & (mag) & (mag) & (mag) & (mag) & (mag) & (mag) & (mag) & (mag)\\
\endfirsthead
\multicolumn{5}{c}{\textbf{Table \ref{tab:magnitudes}} (continued)} \\
\hline
\hline
Star & $B$ & $G_{\rm BP}$ & $V$ & $G$ & $G_{\rm RP}$ & $J$ & $H$ & $K_{S}$ & $I_1$ & $I_2$\\
 & (mag) & (mag) & (mag) & (mag) & (mag) & (mag) & (mag) & (mag) & (mag) & (mag)\\
\hline
\endhead
\hline
\endfoot
\endlastfoot
\hline
V1162 Aql &- &8.06 &7.79 &7.55 &6.92 &6.13 &5.80 &5.69 &-&- \\
TT Aql &- &- &7.13 &6.69 &5.85 &4.66 &4.19 &4.01 &3.88&3.89 \\
U Aql &- &- &6.43 &- &- &4.38 &3.99 &3.85 &3.73&3.73 \\
FM Aql &- &8.60 &8.26 &7.78 &6.90 &5.68 &5.21 &5.02 &-&- \\
SZ Aql &- &- &8.63 &8.11 &7.19 &5.86 &5.34 &5.14 &4.98&4.98 \\
FN Aql &- &8.69 &8.37 &7.92 &7.10 &5.95 &5.49 &5.31 &-&- \\
$\eta$ Aql &- &- &3.87 &- &- &2.40 &2.07 &1.96 &1.86&1.86 \\
SY Aur &- &9.36 &9.05 &8.69 &7.94 &6.92 &6.53 &6.36 &-&- \\
RT Aur &6.11 &5.68 &5.44 &5.32 &4.83 &4.22 &3.98 &3.90 &3.85&3.85 \\
VY Car &- &- &7.46 &7.13 &6.39 &5.37 &4.93 &4.77 &-&- \\
$\ell$ Car &- &- &3.73 &- &- &1.68 &1.21 &1.05 &0.95&1.00 \\
DD Cas &- &10.1 &9.87 &9.45 &8.64 &7.53 &7.08 &6.90 &-&- \\
CF Cas &- &11.4 &11.1 &10.6 &9.79 &8.59 &8.13 &7.95 &7.80&7.79 \\
SW Cas &- &10.0 &9.69 &9.29 &8.50 &7.41 &6.99 &6.81 &-&- \\
DL Cas &- &9.28 &8.96 &8.51 &7.68 &6.54 &6.11 &5.93 &5.79&5.78 \\
KN Cen &- &- &9.85 &9.11 &8.05 &6.41 &5.72 &5.45 &-&- \\
$\delta$ Cep &- &- &3.93 &- &- &2.67 &2.39 &2.29 &2.22&2.21 \\
V0459 Cyg &- &10.9 &10.5 &9.96 &9.00 &7.61 &7.07 &6.85 &-&- \\
SZ Cyg &- &- &9.41 &8.84 &7.91 &6.53 &5.95 &5.73 &-&- \\
V0538 Cyg &- &10.7 &10.4 &9.94 &9.05 &7.81 &7.31 &7.10 &-&- \\
V0402 Cyg &- &10.1 &9.85 &9.50 &8.78 &7.80 &7.41 &7.26 &-&- \\
CD Cyg &- &- &8.95 &8.48 &7.61 &6.36 &5.85 &5.65 &5.49&5.49 \\
X Cyg &- &- &6.39 &6.09 &5.36 &4.38 &3.94 &3.79 &3.69&3.72 \\
MW Cyg &- &9.81 &9.47 &8.93 &8.01 &6.69 &6.20 &5.99 &-&- \\
V0386 Cyg &- &9.94 &9.57 &8.92 &7.88 &6.37 &5.80 &5.54 &-&- \\
VZ Cyg &- &9.22 &8.94 &8.69 &8.04 &7.20 &6.85 &6.72 &-&- \\
$\beta$ Dor &- &- &3.73 &- &- &2.36 &2.03 &1.93 &1.85&1.86 \\
$\zeta$ Gem &- &- &3.88 &- &- &2.52 &2.20 &2.10 &2.02&2.04 \\
V Lac &- &9.22 &8.91 &8.63 &7.95 &7.03 &6.68 &6.53 &-&- \\
BG Lac &- &9.16 &8.87 &8.58 &7.91 &7.02 &6.64 &6.50 &-&- \\
RR Lac &- &9.15 &8.86 &8.56 &7.88 &6.97 &6.62 &6.48 &-&- \\
Z Lac &- &8.75 &8.42 &8.05 &7.28 &6.23 &5.81 &5.65 &-&- \\
Y Lac &- &9.42 &9.15 &8.96 &8.38 &7.63 &7.31 &7.19 &-&- \\
CV Mon &- &10.6 &10.2 &9.68 &8.70 &7.30 &6.78 &6.56 &6.37&6.34 \\
T Mon &- &- &6.13 &5.83 &5.09 &4.09 &3.64 &3.48 &3.38&3.42 \\
S Mus &6.98 &6.40 &6.13 &5.89 &5.26 &4.45 &4.12 &3.99 &-&- \\
S Nor &- &6.71 &6.43 &6.15 &5.50 &4.64 &4.28 &4.15 &4.05&4.05 \\
AW Per &- &7.78 &7.46 &7.08 &6.29 &5.22 &4.84 &4.67 &-&- \\
RS Pup &- &- &7.01 &- &5.63 &4.35 &3.81 &3.60 &-&- \\
AQ Pup &- &- &8.65 &8.17 &7.29 &6.00 &5.46 &5.26 &-&- \\
VZ Pup &10.9 &10.0 &9.64 &9.24 &8.43 &7.27 &6.82 &6.63 &-&- \\
X Pup &- &8.89 &8.50 &8.11 &7.28 &6.09 &5.60 &5.40 &-&- \\
LS Pup &- &10.7 &10.4 &10.0 &9.18 &7.99 &7.51 &7.31 &-&- \\
RY Sco &- &8.37 &8.00 &7.36 &6.38 &4.92 &4.33 &4.10 &-&- \\
V0636 Sco &7.58 &6.92 &6.64 &6.38 &5.73 &4.89 &4.52 &4.39 &-&- \\
SS Sct &- &8.48 &8.20 &7.9 &7.21 &6.29 &5.94 &5.80 &-&- \\
Z Sct &- &- &9.58 &9.09 &8.21 &6.96 &6.48 &6.28 &-&- \\
S Sge &- &5.88 &5.61 &5.43 &4.86 &4.14 &3.84 &3.73 &3.66&3.66 \\
U Sgr &7.80 &7.00 &6.69 &6.31 &5.54 &4.49 &4.09 &3.94 &3.82&3.81 \\
BB Sgr &7.95 &7.24 &6.94 &6.64 &5.94 &5.02 &4.63 &4.50 &-&- \\
XX Sgr &- &9.18 &8.84 &- &7.58 &6.42 &5.97 &5.77 &-&- \\
W Sgr &5.45 &4.92 &4.65 &4.49 &3.93 &3.22 &2.91 &2.80 &2.72&2.72 \\
WZ Sgr &- &- &8.03 &7.52 &6.61 &5.28 &4.75 &4.54 &4.38&4.40 \\
Y Sgr &6.63 &6.01 &5.73 &- &4.85 &4.03 &3.69 &3.57 &3.48&3.48 \\
X Sgr &5.35 &4.80 &4.54 &4.32 &3.72 &2.94 &2.64 &2.52 &2.42&2.40 \\
V0350 Sgr &- &7.76 &7.47 &7.18 &6.51 &5.62 &5.26 &5.12 &-&- \\
ST Tau &9.16 &8.50 &8.20 &7.91 &7.22 &6.30 &5.93 &5.77 &-&- \\
RZ Vel &- &- &7.08 &- &5.97 &4.89 &4.45 &4.27 &-&- \\
U Vul &- &7.43 &7.09 &6.62 &5.75 &4.55 &4.12 &3.93 &3.78&3.77 \\
T Vul &6.39 &5.98 &5.74 &5.62 &5.13 &4.53 &4.27 &4.18 &4.11&4.11 \\
S Vul &- &- &8.95 &- &- &5.43 &4.83 &4.57 &4.35&4.33 \\
SV Vul &- &- &7.21 &6.72 &5.84 &4.58 &4.07 &3.87 &3.73&3.76 \\
X Vul &- &9.17 &8.81 &8.25 &7.29 &5.93 &5.43 &5.21 &-&- \\
\hline
\label{tab:magnitudes}
\end{longtable} 
\tablefoot{The statistical uncertainties are up to 0.008 mag, and we considered a 0.01 systematic uncertainty in order to take the photometric zeropoints into account. For some Cepheids, $B$ photometry was available, but was not taken into account in the modeling because the temperature of the star is low (see Sect. \ref{sec:spips}))}
\clearpage
\newpage

\section{Period changes}
\setlength{\tabcolsep}{6pt}
{\footnotesize
\begin{longtable}{l c c c c c c c c c}
\caption{Period changes dP/dt (in s/yr) derived from the SPIPS models.} \\
\hline
\hline
Star & MJD0 &dP/dt&P$_0$ &P$_1$ & P$_2$ & P$_3$ & P$_4$ & P$_5$ & P$_6$\\
\endfirsthead
\multicolumn{9}{c}{\textbf{Table \ref{tab:periods}} (continued)} \\
\hline
\hline
Star & MJD0 &dP/dt &P$_0$ & P$_1$ & P$_2$ & P$_3$ & P$_4$ & P$_5$ & P$_6$\\
\hline 
\endhead
\hline
\endfoot
\endlastfoot
\hline
V1162 Aql       &       25802.823 &      0.077&5.376$_{\pm 6.10^{-06}}$& 2.10$^{-09}_{\pm 1.10^{-10}}$&  -&      -&      -&      -&      -               \\
TT Aql  &       48308.571 &      -0.549&13.755$_{\pm 3.10^{-05}}$& -2.10$^{-08}_{\pm 5.10^{-09}}$&   -&      -&      -&      -&      -               \\
U Aql   &       34922.086 &      0.235&7.024$_{\pm 2.10^{-05}}$& 7.10$^{-09}_{\pm 6.10^{-10}}$&   -&      -&      -&      -&      -               \\
FM Aql  &       35151.203 &      -0.022&6.114$_{\pm 7.10^{-06}}$& -7.10$^{-10}_{\pm 2.10^{-10}}$&   -&      -&      -&      -&      -               \\
SZ Aql  &       54228.333 &      3.821&17.142$_{\pm 6.10^{-05}}$& 1.10$^{-07}_{\pm 6.10^{-09}}$&   -&      -&      -&      -&      -               \\
FN Aql  &       36803.278 &      -1.757&9.483$_{\pm 4.10^{-05}}$& -6.10$^{-08}_{\pm 1.10^{-09}}$&   -&      -&      -&      -&      -               \\
$\eta$ Aql      &       48069.390 &      -0.092&7.177$_{\pm 1.10^{-05}}$& -3.10$^{-09}_{\pm 2.10^{-09}}$& -&      -&      -&      -&      -               \\
SY Aur  &       36843.274 &      0.989&10.144$_{\pm 6.10^{-05}}$& 3.10$^{-08}_{\pm 2.10^{-09}}$&   -&      -&      -&      -&      -               \\
RT Aur  &       47956.905 &      -0.196&3.728$_{\pm 5.10^{-06}}$& -6.10$^{-09}_{\pm 6.10^{-10}}$&   -&      -&      -&      -&      -               \\
VY Car  &       48339.297 &      -36.269&18.902$_{\pm 4.10^{-05}}$& -1.10$^{-06}_{\pm 6.10^{-09}}$&   -&      -&      -&      -&      -               \\
$\ell$ Car      &       47774.237 &      27.558&35.552$_{\pm 1.10^{-03}}$& 1.10$^{-06}_{\pm 2.10^{-07}}$&  -6.10$^{-12}_{\pm 2.10^{-11}}$& -4.10$^{-15}_{\pm 3.10^{-15}}$&   -&      -&      -               \\
DD Cas  &       42780.178 &      1.240&9.811$_{\pm 3.10^{-05}}$& 4.10$^{-08}_{\pm 2.10^{-09}}$&   -&      -&      -&      -&      -               \\
CF Cas  &       37021.259 &      -0.344&4.875$_{\pm 6.10^{-06}}$& -1.10$^{-08}_{\pm 2.10^{-10}}$&   -&      -&      -&      -&      -               \\
SW Cas  &       42989.081 &      -0.316&5.441$_{\pm 1.10^{-05}}$& -1.10$^{-08}_{\pm 8.10^{-10}}$&   -&      -&      -&      -&      -               \\
DL Cas  &       42779.729 &      -0.040&8.001$_{\pm 2.10^{-05}}$& -1.10$^{-09}_{\pm 1.10^{-09}}$&   -&      -&      -&      -&      -               \\
KN Cen  &       54345.370 &      -34.352&34.019$_{\pm 2.10^{-04}}$& -1.10$^{-06}_{\pm 2.10^{-08}}$&   -&      -&      -&      -&      -               \\
$\delta$ Cep    &       36075.009 &      -0.013&5.366$_{\pm 2.10^{-06}}$& -4.10$^{-10}_{\pm 6.10^{-11}}$& -&      -&      -&      -&      -               \\
V0459 Cyg       &       36807.804 &      -0.258&7.252$_{\pm 2.10^{-05}}$& -8.10$^{-09}_{\pm 9.10^{-10}}$& -&      -&      -&      -&      -               \\
SZ Cyg  &       43306.955 &      0.890&15.109$_{\pm 1.10^{-04}}$& 3.10$^{-08}_{\pm 7.10^{-09}}$&   -&      -&      -&      -&      -               \\
V0538 Cyg       &       42772.448 &      -0.025&6.119$_{\pm 3.10^{-05}}$& -8.10$^{-10}_{\pm 2.10^{-09}}$& -&      -&      -&      -&      -               \\
V0402 Cyg       &       41698.052 &      -0.295&4.365$_{\pm 7.10^{-06}}$& -9.10$^{-09}_{\pm 3.10^{-10}}$& -&      -&      -&      -&      -               \\
CD Cyg  &       48321.640 &      0.676&17.075$_{\pm 6.10^{-05}}$& 2.10$^{-08}_{\pm 9.10^{-09}}$&   -&      -&      -&      -&      -               \\
X Cyg   &       48319.538 &      1.709&16.386$_{\pm 3.10^{-05}}$& 5.10$^{-08}_{\pm 4.10^{-09}}$&   -&      -&      -&      -&      -               \\
MW Cyg  &       42923.409 &      0.099&5.955$_{\pm 1.10^{-05}}$& 3.10$^{-09}_{\pm 8.10^{-10}}$&   -&      -&      -&      -&      -               \\
V0386 Cyg       &       42776.457 &      -0.556&5.258$_{\pm 2.10^{-05}}$& -2.10$^{-08}_{\pm 8.10^{-10}}$& -&      -&      -&      -&      -               \\
VZ Cyg  &       41705.189 &      -0.188&4.864$_{\pm 5.10^{-06}}$& -6.10$^{-09}_{\pm 3.10^{-10}}$&   -&      -&      -&      -&      -               \\
$\beta$ Dor     &       50274.946 &      -0.060&9.843$_{\pm 2.10^{-05}}$& -2.10$^{-09}_{\pm 3.10^{-09}}$& -&      -&      -&      -&      -               \\
$\zeta$ Gem     &       48707.923 &      -0.929&10.150$_{\pm 3.10^{-05}}$& -3.10$^{-08}_{\pm 7.10^{-09}}$& -&      -&      -&      -&      -               \\
V Lac   &       28900.559 &      -0.510&4.984$_{\pm 3.10^{-06}}$& -2.10$^{-08}_{\pm 1.10^{-10}}$&   -&      -&      -&      -&      -               \\
BG Lac  &       35314.337 &      -0.255&5.332$_{\pm 6.10^{-06}}$& -8.10$^{-09}_{\pm 2.10^{-10}}$&   -&      -&      -&      -&      -               \\
RR Lac  &       42776.203 &      -0.150&6.416$_{\pm 1.10^{-05}}$& -5.10$^{-09}_{\pm 5.10^{-10}}$&   -&      -&      -&      -&      -               \\
Z Lac   &       48313.070 &      0.270&10.886$_{\pm 3.10^{-05}}$& 9.10$^{-09}_{\pm 3.10^{-09}}$&   -&      -&      -&      -&      -               \\
Y Lac   &       41746.264 &      -0.004&4.324$_{\pm 3.10^{-06}}$& -1.10$^{-10}_{\pm 2.10^{-10}}$&   -&      -&      -&      -&      -               \\
CV Mon  &       42772.649 &      0.055&5.379$_{\pm 1.10^{-05}}$& 2.10$^{-09}_{\pm 8.10^{-10}}$&   -&      -&      -&      -&      -               \\
T Mon   &       43783.953 &      15.945&27.026$_{\pm 3.10^{-04}}$& 5.10$^{-07}_{\pm 3.10^{-08}}$&   -&      -&      -&      -&      -               \\
S Mus   &       40300.762 &      0.224&9.660$_{\pm 3.10^{-05}}$& 7.10$^{-09}_{\pm 1.10^{-09}}$&   -&      -&      -&      -&      -               \\
S Nor   &       44018.558 &      0.412&9.754$_{\pm 3.10^{-05}}$& 1.10$^{-08}_{\pm 2.10^{-09}}$&   -&      -&      -&      -&      -               \\
AW Per  &       42708.656 &      0.158&6.464$_{\pm 1.10^{-05}}$& 5.10$^{-09}_{\pm 9.10^{-10}}$&   -&      -&      -&      -&      -               \\
RS Pup  &       54215.800 &      23.042&41.454$_{\pm 7.10^{-04}}$& -5.10$^{-06}_{\pm 2.10^{-07}}$&   5.10$^{-09}_{\pm 6.10^{-11}}$&  1.10$^{-12}_{\pm 7.10^{-15}}$&  1.10$^{-16}_{\pm 4.10^{-19}}$&   3.10$^{-21}_{\pm 2.10^{-23}}$&  -               \\
AQ Pup  &       54587.136 &      133.297&30.167$_{\pm 3.10^{-04}}$& 4.10$^{-06}_{\pm 8.10^{-08}}$&   3.10$^{-11}_{\pm 8.10^{-12}}$&  -&      -&      -&      -               \\
VZ Pup  &       41121.154 &      2.934&23.174$_{\pm 2.10^{-04}}$& 9.10$^{-08}_{\pm 8.10^{-09}}$&   -&      -&      -&      -&      -               \\
X Pup   &       54143.669 &      7.752&25.971$_{\pm 1.10^{-04}}$& 1.10$^{-07}_{\pm 6.10^{-08}}$&   -2.10$^{-11}_{\pm 5.10^{-12}}$& -&      -&      -&      -               \\
LS Pup  &       38375.646 &      -0.009&14.147$_{\pm 6.10^{-05}}$& -3.10$^{-10}_{\pm 2.10^{-09}}$&   -&      -&      -&      -&      -               \\
RY Sco  &       54670.502 &      4.005&20.323$_{\pm 2.10^{-04}}$& 2.10$^{-07}_{\pm 7.10^{-08}}$&   7.10$^{-12}_{\pm 4.10^{-12}}$&  -&      -&      -&      -               \\
V0636 Sco       &       51402.316 &      -0.052&6.797$_{\pm 8.10^{-06}}$& -2.10$^{-09}_{\pm 1.10^{-09}}$& -&      -&      -&      -&      -               \\
SS Sct  &       35315.072 &      0.026&3.671$_{\pm 3.10^{-06}}$& 8.10$^{-10}_{\pm 1.10^{-10}}$&   -&      -&      -&      -&      -               \\
Z Sct   &       36246.638 &      0.800&12.901$_{\pm 5.10^{-05}}$& 3.10$^{-08}_{\pm 2.10^{-09}}$&   -&      -&      -&      -&      -               \\
S Sge   &       42678.306 &      -0.003&8.382$_{\pm 2.10^{-05}}$& -1.10$^{-10}_{\pm 1.10^{-09}}$&   -&      -&      -&      -&      -               \\
U Sgr   &       30117.481 &      -0.039&6.745$_{\pm 6.10^{-06}}$& -1.10$^{-09}_{\pm 2.10^{-10}}$&   -&      -&      -&      -&      -               \\
BB Sgr  &       36053.022 &      0.090&6.637$_{\pm 1.10^{-05}}$& 3.10$^{-09}_{\pm 5.10^{-10}}$&   -&      -&      -&      -&      -               \\
XX Sgr  &       52839.717 &      -0.069&6.424$_{\pm 1.10^{-05}}$& -2.10$^{-09}_{\pm 2.10^{-09}}$&   -&      -&      -&      -&      -               \\
W Sgr   &       48690.679 &      0.145&7.595$_{\pm 1.10^{-05}}$& 5.10$^{-09}_{\pm 1.10^{-09}}$&   -2.10$^{-13}_{\pm 2.10^{-13}}$& -&      -&      -&      -               \\
WZ Sgr  &       35506.573 &      4.385&21.848$_{\pm 1.10^{-04}}$& 1.10$^{-07}_{\pm 4.10^{-09}}$&   -&      -&      -&      -&      -               \\
Y Sgr   &       47303.128 &      0.022&5.773$_{\pm 7.10^{-06}}$& 7.10$^{-10}_{\pm 9.10^{-10}}$&   -&      -&      -&      -&      -               \\
X Sgr   &       48707.915 &      0.047&7.013$_{\pm 1.10^{-05}}$& 1.10$^{-09}_{\pm 2.10^{-09}}$&   -&      -&      -&      -&      -               \\
V0350 Sgr       &       35316.260 &      -0.215&5.154$_{\pm 7.10^{-06}}$& -7.10$^{-09}_{\pm 2.10^{-10}}$& -&      -&      -&      -&      -               \\
ST Tau  &       41761.544 &      0.077&4.034$_{\pm 3.10^{-06}}$& 2.10$^{-09}_{\pm 2.10^{-10}}$&   -&      -&      -&      -&      -               \\
RZ Vel  &       34845.924 &      3.658&20.395$_{\pm 5.10^{-05}}$& 1.10$^{-07}_{\pm 2.10^{-09}}$&   -&      -&      -&      -&      -               \\
U Vul   &       48311.104 &      -0.161&7.991$_{\pm 2.10^{-05}}$& -5.10$^{-09}_{\pm 3.10^{-09}}$&   -&      -&      -&      -&      -               \\
T Vul   &       41704.726 &      -0.077&4.435$_{\pm 2.10^{-06}}$& -2.10$^{-08}_{\pm 2.10^{-09}}$&   1.10$^{-12}_{\pm 3.10^{-13}}$&  -3.10$^{-17}_{\pm 1.10^{-17}}$& -&      -&      -               \\
S Vul   &       48332.000 &      -840.466&68.552$_{\pm 6.10^{-03}}$& -9.10$^{-05}_{\pm 3.10^{-06}}$&   -4.10$^{-09}_{\pm 8.10^{-10}}$& 2.10$^{-12}_{\pm 1.10^{-13}}$&  -9.10$^{-17}_{\pm 3.10^{-17}}$&   -8.10$^{-22}_{\pm 1.10^{-21}}$& 4.10$^{-26}_{\pm 4.10^{-26}}$           \\
SV Vul  &       48307.758 &      -248.143&44.941$_{\pm 1.10^{-03}}$& -3.10$^{-05}_{\pm 4.10^{-07}}$&   4.10$^{-09}_{\pm 1.10^{-10}}$&  6.10$^{-13}_{\pm 2.10^{-14}}$&  -1.10$^{-16}_{\pm 4.10^{-18}}$&   2.10$^{-21}_{\pm 3.10^{-22}}$&  1.10$^{-25}_{\pm 7.10^{-27}}$           \\
X Vul   &       35308.510 &      -0.660&6.320$_{\pm 9.10^{-06}}$& -2.10$^{-08}_{\pm 3.10^{-10}}$&   -&      -&      -&      -&      -               \\
\hline
\label{tab:periods}
\end{longtable} 
\tablefoot{The period P(x) in days is given by the polynomial expression $P(x) = P_0 + P_1 x + P_2x^2 + P_3 x^3 + P_4x^4 + P_5x^5 + P_6x^6$ with $x = \rm MJD - \rm MJD_0$ and $P_0$  in days, $P_1$ in s/year.}
}
\clearpage
\newpage

\section{Fourier coefficients}
\begin{table}[h!]
\centering
\footnotesize
\caption{First three order coefficients of the Fourier series in the $K$ -band photometry (not dereddened). Fourier series are of the form $f(x) = A_0 + \sum^N_{i=1}{A_i\cos(2\pi ix + \phi_i)}$.}
\begin{tabular}{l c c c c c c c}
\hline
\hline
Star & $A_0$ &$A_1$ & $A_2$ & $A_3$ & $\phi_1$ & $\phi_2$ & $\phi_3$ \\
\hline
V1162 Aql       & 5.71& 0.08& -0.02& -0.01& 1.48& -0.41 &0.81   \\
TT Aql  & 4.04& 0.16& -0.02& -0.01& 1.18& -0.91 &-0.43  \\
U Aql   & 3.87& 0.10& -0.03& -0.01& 1.35& -0.60 &0.41   \\
FM Aql  & 5.04& 0.08& -0.02& -0.01& 1.58& -0.45 &0.26   \\
SZ Aql  & 5.17& 0.17& 0.03& -0.01& 1.05& 1.86   &-0.55  \\
FN Aql  & 5.34& 0.11& 0.01& 0.00& 1.92& -1.37   &0.00   \\
$\eta$ Aql      & 1.98& 0.10& -0.02& -0.01& 1.40& -0.45 &0.32   \\
SY Aur  & 6.38& -0.08& -0.02& -0.00& -1.01& 0.91        &0.19   \\
RT Aur  & 3.93& 0.08& -0.02& -0.01& 1.60& -0.61 &0.05   \\
VY Car  & 4.80& 0.18& -0.03& -0.01& 1.10& -1.32 &-0.77  \\
$\ell$ Car      & 1.08& 0.15& -0.03& -0.01& 1.17& -1.15 &-0.30  \\
DD Cas  & 6.92& 0.11& -0.01& -0.00& 1.66& 1.23  &-5.06  \\
CF Cas  & 7.97& 0.08& -0.02& -0.01& 1.48& -0.53 &0.53   \\
SW Cas  & 6.83& 0.09& -0.03& -0.01& 1.54& -0.50 &0.47   \\
DL Cas  & 5.95& 0.08& -0.02& -0.01& 1.44& -0.45 &0.82   \\
KN Cen  & 5.48& 0.19& 0.04& -0.02& 1.01& -4.51  &-0.67  \\
$\delta$ Cep    & 2.32& 0.09& -0.03& -0.01& 1.51& -0.61 &0.18   \\
V0459 Cyg       & 6.88& 0.09& -0.02& -0.01& 1.37& -0.55 &0.51   \\
SZ Cyg  & 5.75& 0.16& -0.02& -0.01& 1.30& -0.97 &-0.39  \\
V0538 Cyg       & 7.13& 0.08& -0.02& -0.00& 1.50& -0.37 &0.59   \\
V0402 Cyg       & 7.29& 0.08& -0.02& -0.01& 1.68& -0.29 &0.66   \\
CD Cyg  & 5.68& 0.18& 0.03& -0.01& 0.98& -4.46  &-0.59  \\
X Cyg   & 3.82& 0.17& -0.03& -0.01& 1.15& 5.04  &-0.75  \\
MW Cyg  & 6.02& 0.09& -0.03& -0.01& 1.37& -0.60 &0.14   \\
V0386 Cyg       & 5.57& 0.09& -0.02& -0.01& 1.44& -0.76 &0.00   \\
VZ Cyg  & 6.75& 0.09& -0.02& -0.01& 1.44& -0.68 &0.06   \\
$\beta$ Dor     & 1.95& 0.10& -0.01& -0.00& 1.81& 1.29  &1.03   \\
$\zeta$ Gem     & 2.12& 0.09& 0.01& 0.00& 1.91& -1.49   &-0.94  \\
V Lac   & 6.56& 0.10& -0.03& -0.02& 1.51& -0.71 &-0.03  \\
BG Lac  & 6.53& 0.08& -0.02& -0.01& 1.53& -0.47 &0.44   \\
RR Lac  & 6.51& 0.09& -0.03& -0.01& 1.63& -0.39 &0.28   \\
Z Lac   & 5.67& 0.13& -0.02& -0.01& 1.60& 0.53  &0.40   \\
Y Lac   & 7.21& 0.08& -0.03& -0.01& 1.60& -0.48 &0.29   \\
CV Mon  & 6.58& 0.09& -0.03& -0.01& 1.41& -0.70 &0.08   \\
T Mon   & 3.51& 0.18& -0.04& -0.02& 1.18& -1.13 &-0.44  \\
S Mus   & 4.01& -0.08& -0.02& 0.00& -0.81& -4.50        &-4.03  \\
S Nor   & 4.18& -0.09& -0.01& 0.00& -1.09& -4.58        &0.00   \\
AW Per  & 4.70& 0.09& -0.03& -0.02& 1.47& -0.57 &0.05   \\
RS Pup  & 3.63& 0.19& 0.03& -0.01& 1.00& -4.48  &-0.59  \\
AQ Pup  & 5.30& 0.22& 0.05& -0.02& 0.96& -4.52  &-0.75  \\
VZ Pup  & 6.66& 0.22& -0.05& -0.03& 1.00& -1.22 &-0.29  \\
X Pup   & 5.43& 0.18& -0.05& -0.02& 0.94& 4.77  &-0.74  \\
LS Pup  & 7.34& 0.16& -0.02& -0.01& 1.28& -0.31 &-0.06  \\
RY Sco  & 4.13& 0.13& -0.02& -0.01& 1.28& -0.63 &0.31   \\
V0636 Sco       & 4.42& 0.08& -0.02& -0.00& 1.53& -0.24 &1.02   \\
SS Sct  & 5.83& 0.06& -0.02& -0.01& 1.71& -0.48 &0.57   \\
Z Sct   & 6.31& 0.14& -0.02& -0.01& 1.41& -0.02 &0.06   \\
S Sge   & 3.76& 0.09& -0.02& -0.01& 1.31& -0.69 &0.98   \\
U Sgr   & 3.96& 0.09& -0.02& -0.01& 1.42& -0.53 &0.28   \\
BB Sgr  & 4.52& 0.08& -0.02& -0.01& 1.48& -0.28 &0.69   \\
XX Sgr  & 5.80& 0.11& -0.03& -0.02& 1.35& -0.80 &-0.13  \\
W Sgr   & 2.83& 0.10& -0.03& -0.02& 1.28& -0.79 &0.42   \\
WZ Sgr  & 4.57& 0.18& -0.04& -0.02& 1.11& -1.23 &-0.64  \\
Y Sgr   & 3.60& 0.09& -0.02& -0.01& 1.55& -0.49 &0.23   \\
X Sgr   & 2.55& 0.08& -0.02& -0.01& 1.83& 0.05  &0.83   \\
V0350 Sgr       & 5.15& 0.09& -0.03& -0.01& 1.47& -0.59 &0.22   \\
ST Tau  & 5.80& 0.09& -0.03& -0.01& 1.56& -0.68 &-0.07  \\
RZ Vel  & 4.30& 0.18& 0.04& -0.02& 0.92& -4.62  &-0.81  \\
U Vul   & 3.96& 0.08& -0.02& -0.01& 1.48& -0.74 &0.65   \\
T Vul   & 4.21& 0.09& -0.02& -0.01& 1.63& -0.41 &0.40   \\
S Vul   & 4.59& 0.12& 0.01& -0.00& 0.96& -4.49  &-0.49  \\
SV Vul  & 3.90& 0.18& -0.03& -0.01& 0.78& 10.90 &-0.67  \\
X Vul   & 5.23& 0.10& -0.03& -0.01& 1.39& -0.69 &-0.22  \\
\hline
\label{tab:fourier}
\end{tabular}
\end{table} 

\newpage




\section{Dependence of the projection factor on other parameters}

\begin{figure*}[h!]
\centering
\includegraphics[width=\textwidth]{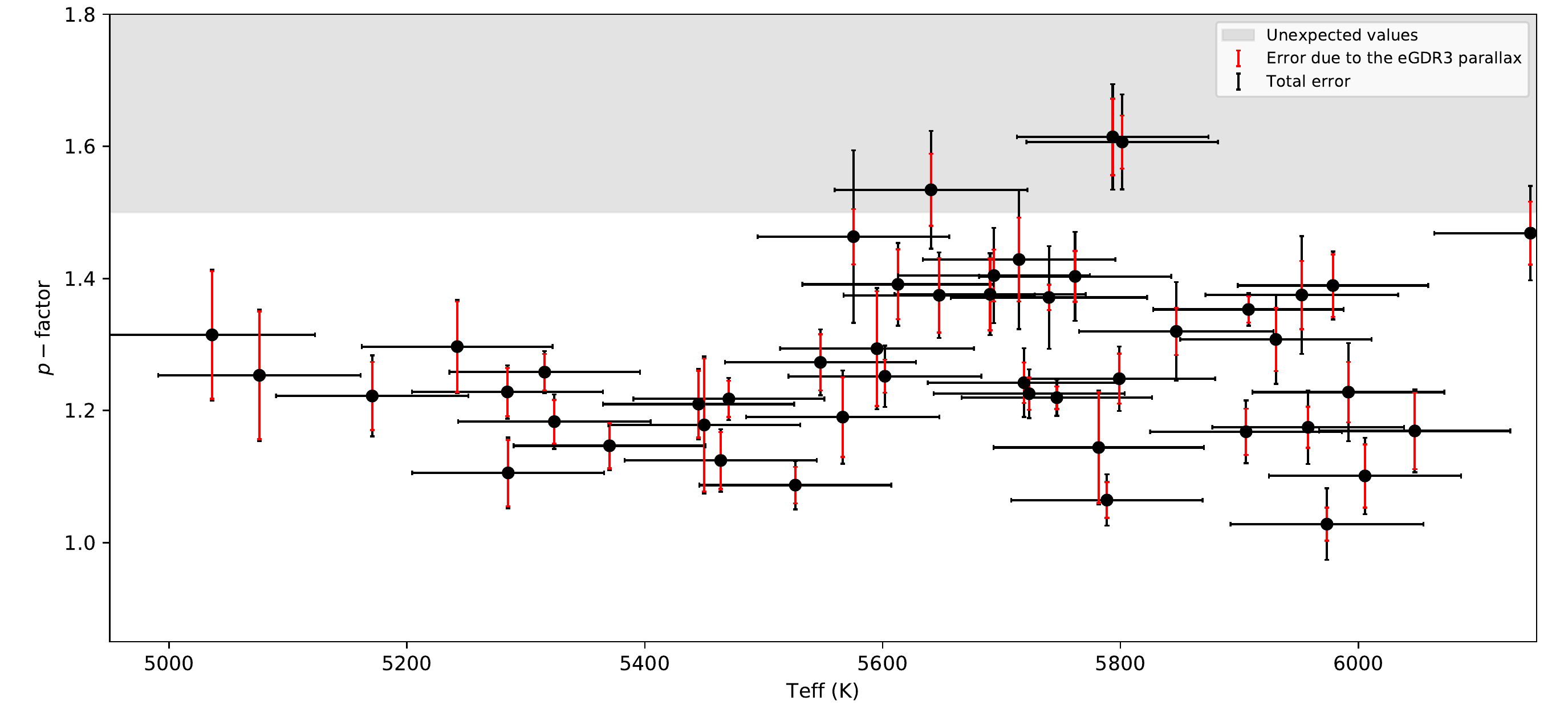}
\caption{Projection factor as a function of the effective temperature (RUWE<1.4 only).}
\label{fig:Teffp}
\end{figure*}

\begin{figure*}[h!]
\centering
\includegraphics[width=\textwidth]{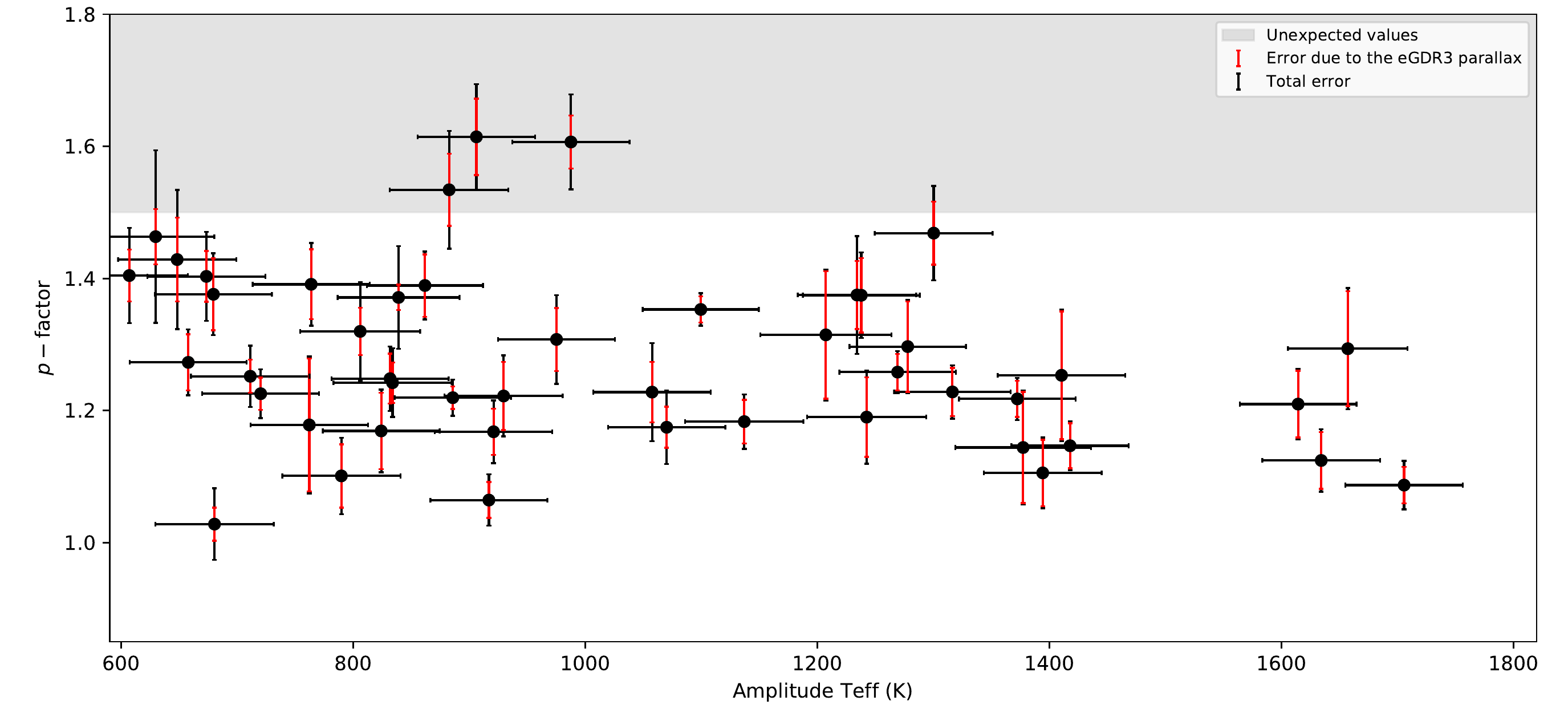}
\caption{Projection factor as a function of the amplitude of the effective temperature (RUWE<1.4 only).}
\label{fig:ampTeffp}
\end{figure*}

\begin{figure*}[h!]
\centering
\includegraphics[width=\textwidth]{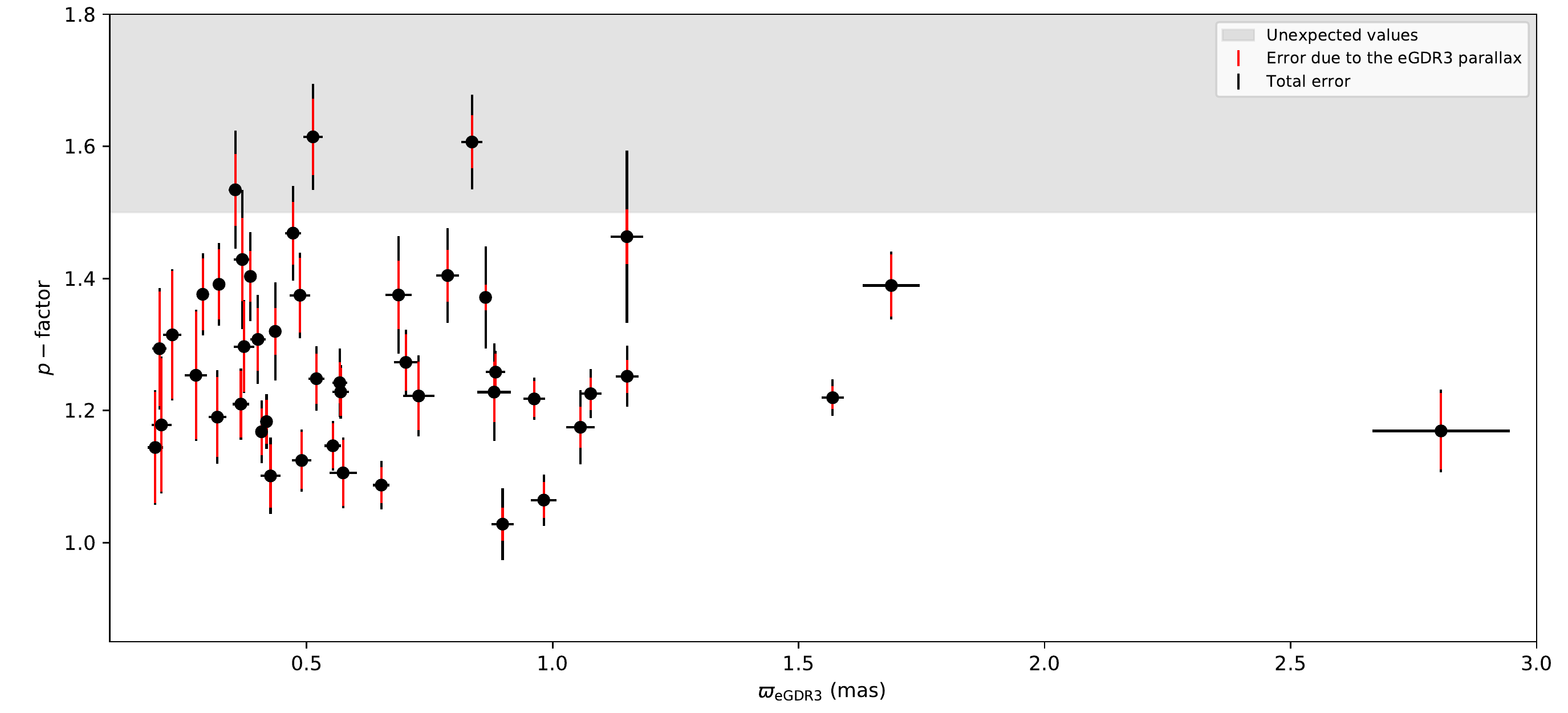}
\caption{Projection factor as a function of the \textit{Gaia} EDR3 parallax (RUWE<1.4 only).}
\label{fig:plxp}
\end{figure*}

\begin{figure*}[h!]
\centering
\includegraphics[width=\textwidth]{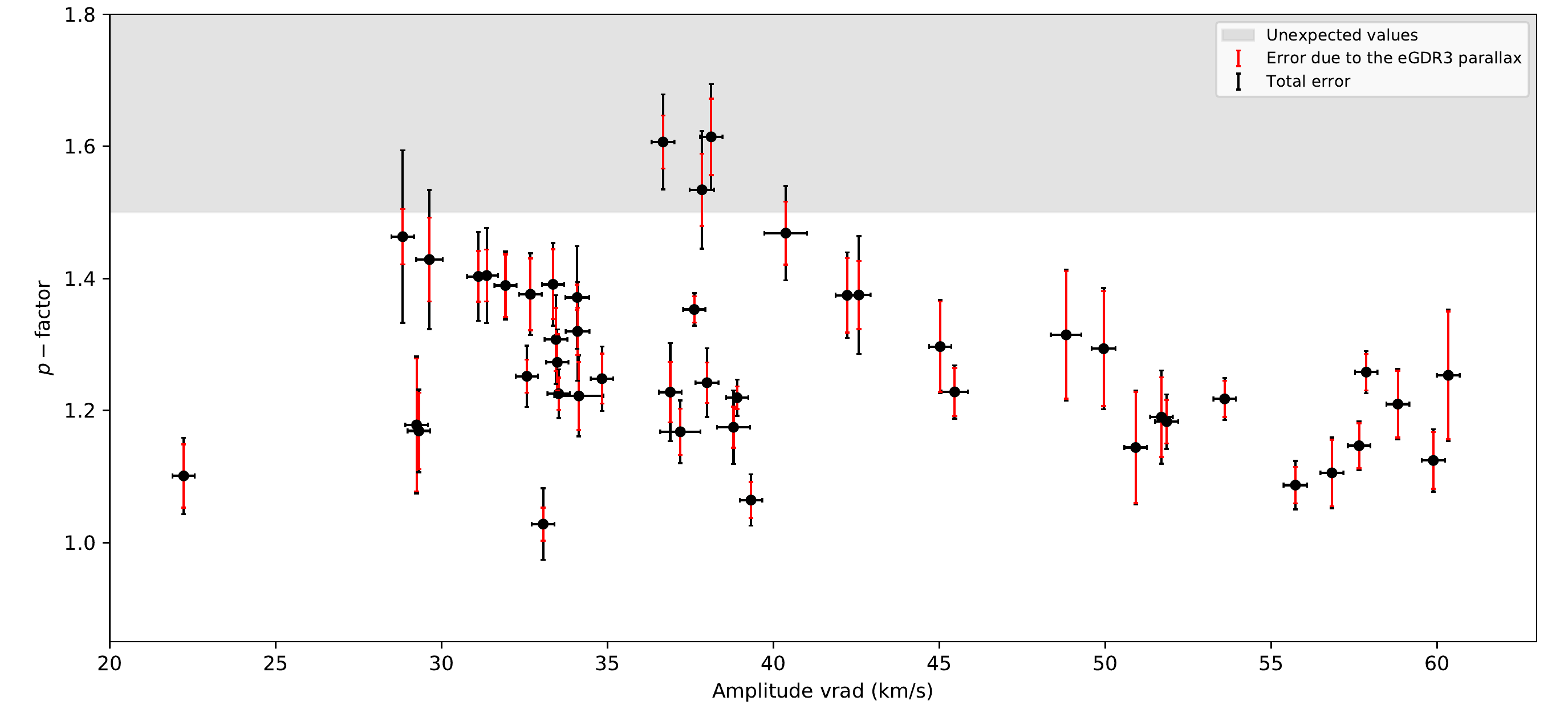}
\caption{Projection factor as a function of the radial velocity amplitude (RUWE<1.4 only).}
\label{fig:dvradp}
\end{figure*}

\begin{figure*}[h!]
\centering
\includegraphics[width=\textwidth]{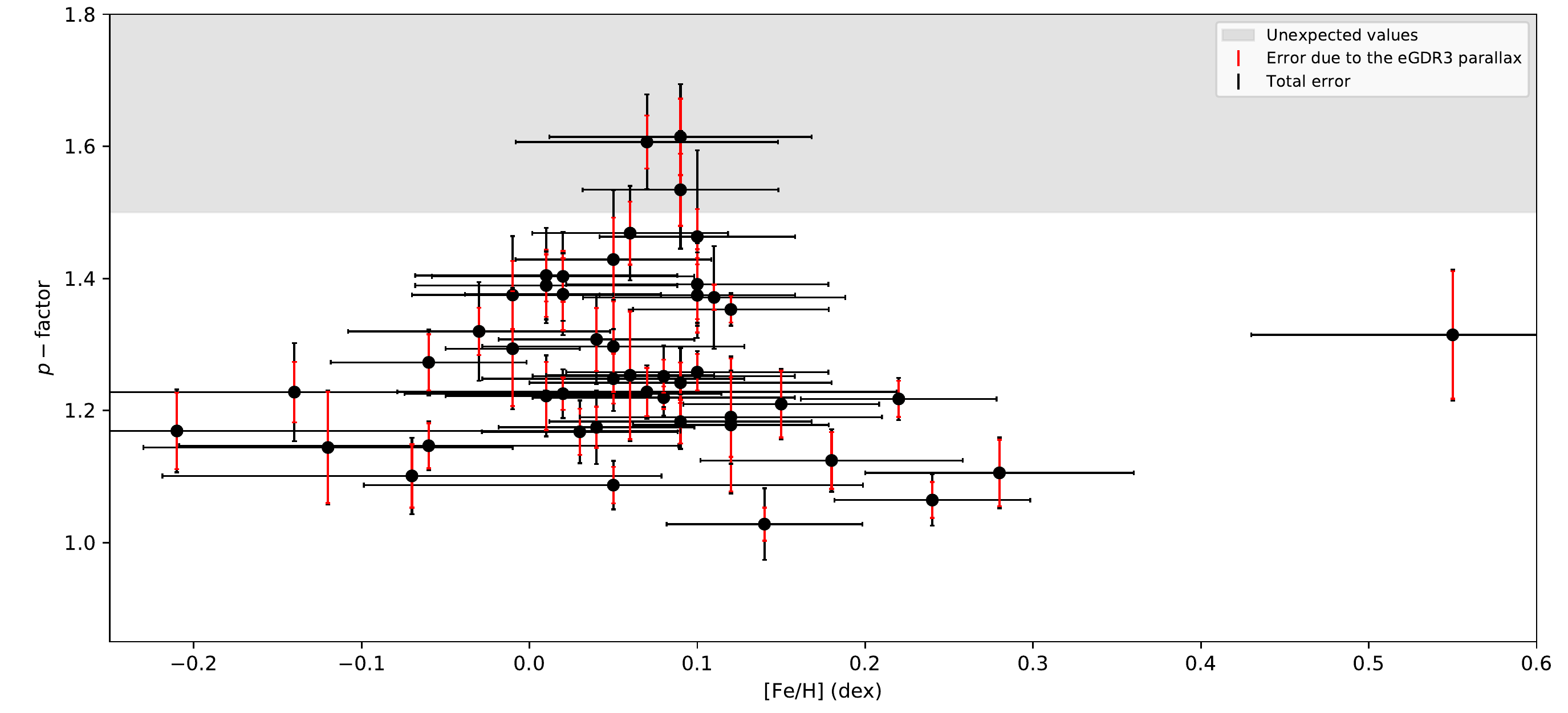}
\caption{Projection factor as a function of the metallicity \citep[taken from][]{Genovali2014, Genovali2015} (RUWE<1.4 only).}
\label{fig:metalp}
\end{figure*}

\begin{figure*}[h!]
\centering
\includegraphics[width=\textwidth]{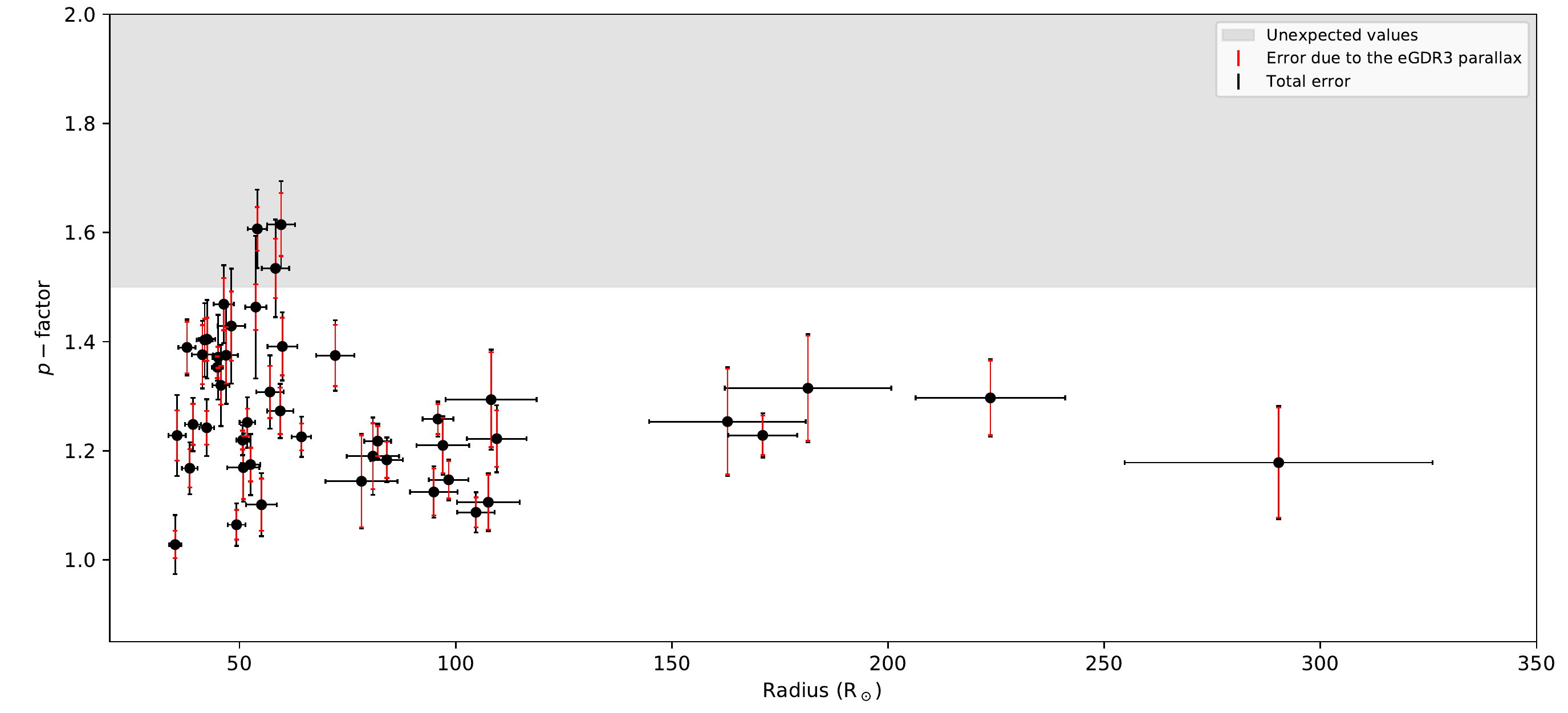}
\caption{Projection factor as a function of the radius (RUWE<1.4 only).}
\label{fig:radiusp}
\end{figure*}

\begin{figure*}[h!]
\centering
\includegraphics[width=\textwidth]{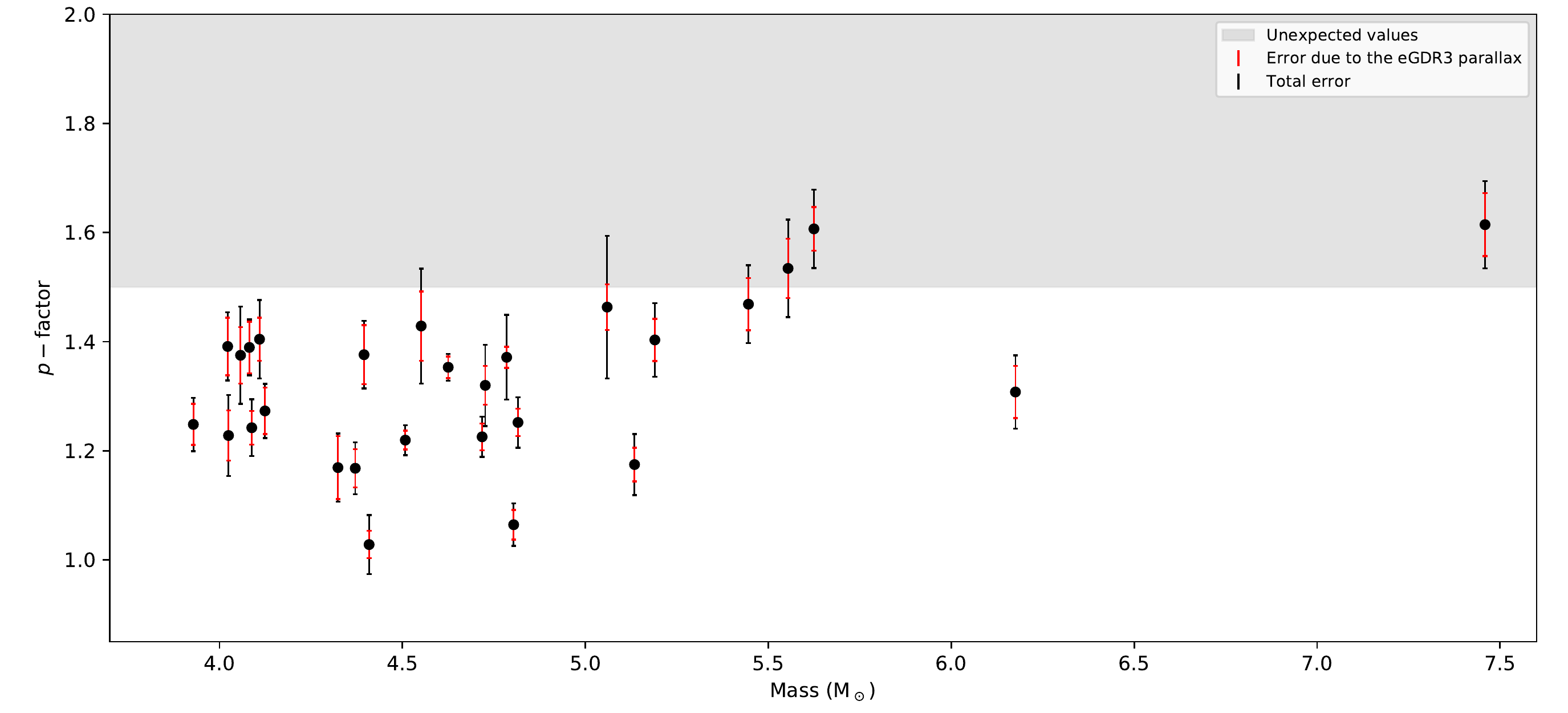}
\caption{Projection factor as a function of the mass (RUWE<1.4 and P<10 days only). The masses were derived using the period-mass-radius relation by \cite{Pilecki2018}, which is applicable up to P=10 days.}
\label{fig:massp}
\end{figure*}

\end{appendix}

\end{document}